\documentclass{article}

\usepackage{jcappub} 
\usepackage[utf8]{inputenc}
\usepackage{appendix}
\usepackage{lineno}

\usepackage{subfigure}
\usepackage[sort&compress]{natbib}
\usepackage{physics}

\title{On the transition radiation interpretation of anomalous ANITA events}

\author{Juan Ammerman-Yebra,}
\author{Jaime Alvarez-Mu\~niz,}
\author{Enrique Zas}

\emailAdd{juan.ammerman.yebra@usc.es}
\emailAdd{jaime.alvarez@usc.es}
\emailAdd{enrique.zas@usc.es}

\affiliation{Instituto Galego de F\'\i sica de Altas Enerx\'\i as (IGFAE),
Universidade de Santiago de Compostela, 15782 Santiago
de Compostela, Spain}

\date{\today}

\begin{document}
\abstract{
The Antarctic Impulsive Transient Antenna (ANITA) detector has observed several radio pulses coming from the surface of the ice cap at the South Pole. These pulses were attributed to upward-going atmospheric particle showers instead of the downward-going showers induced by cosmic rays that exhibit a characteristic polarity inversion of the radio signal due to reflection in the ice. Coherent transition radiation from cosmic-ray showers developing in the atmosphere and intercepting the ice surface has been suggested as a possible and alternative explanation of these so-called ``anomalous'' events. To test this interpretation, we have developed an extension of ZHS, a program to calculate coherent pulses from electromagnetic showers, to deal with showers that transit a planar interface between two homogeneous and dielectric media, including transition radiation. By considering different geometries, it is found that all pulses from air showers intercepting the ice surface and detected at the height of ANITA, display the same polarity as pulses emitted by ultra-high-energy cosmic-ray showers that fully develop in the atmosphere and are reflected on the ice. We find that transition radiation is disfavored as a possible explanation of the anomalous ANITA events. 
}

\maketitle

\section{Introduction}

The Antarctic Impulsive Transient Antenna (ANITA) project \cite{ANITA:2008mzi} was conceived to search for coherent radio pulses from Ultra High Energy (UHE) showers developing in the Antarctic ice sheet, with the main aim of detecting UHE neutrinos. 
Four instruments have been flown in super-pressure balloons at 30-39 km altitudes over the Antarctic continent. While no radio signals due to UHE neutrino interactions in the ice have yet been observed in any of these flights, pulses consistent with coherent radio emission from UHE cosmic ray (UHECR) air showers were detected in the ANITA I flight~\cite{hoover2010observation, ANITA:2016vrp} and in subsequent flights III \cite{ANITA:2018sgj} and IV \cite{ANITA:2020gmv}. The geomagnetic effect dominates the radio emission from air showers~\cite{allan:1969} with a characteristic polarization that has a large component parallel to the Earth surface because the magnetic field is nearly vertical in Antarctica. 
A fraction of the observed pulses comes from stratospheric air showers that develop at high altitudes with tangential trajectories that do not intercept the Earth's surface and point directly to the ANITA detector \cite{Tueros:2023ifq,Tueros:2023lva}, also referred to as ``direct events''. However, most of the detected pulses are reflected on the ice before reaching the detector, and display a characteristic polarity inversion relative to direct events. The rate of these reflected events has been shown to be consistent with the UHECR spectrum~\cite{Schoorlemmer:2015afa}.

In addition, a number of ``anomalous'' pulses have been reported coming from directions below the horizon in the ANITA I~\cite{ANITA:2016vrp}, III~\cite{ANITA:2018sgj} and IV~\cite{ANITA:2020gmv} flights.
They are horizontally polarized, as expected for air showers, but they do not display the characteristic polarity inversion expected from reflection of UHECR-induced pulses. 
The first two such anomalous events correspond to elevation angles of $27.4^\circ$ and $35.0^\circ$ as measured respectively at the ANITA I~\cite{ANITA:2016vrp} and III~\cite{ANITA:2018sgj} instruments, i.e. they travelled in the upward-direction with respect to the ground. 
Upward-going showers are expected to be induced in the atmosphere by the decay of leptons, particularly 
tau leptons that can traverse larger amounts of matter before decaying \cite{Bertou:2001vm,Fargion:2000iz}. Tau leptons are produced in charged current tau neutrino interactions. Tau neutrinos can traverse through most of the Earth and interact below the surface so that a tau lepton exits into the atmosphere in the upward direction. The energy of these neutrinos must be larger than $\sim 100$ PeV for the pulses induced by the tau decay showers to be detected by the ANITA instruments. Early reports of the anomalous events already pointed out that the elevation angle of the observed pulses was too high for UHE neutrinos to be able to traverse thousands of kilometers through the interior of Earth before interacting just below the ice~\cite{ANITA:2016vrp}. 
The tau neutrino interpretation of these events 
has in fact been ruled out by detailed simulation of the pulses and a comparison of the implied neutrino fluxes with upper bounds from experiments such as IceCube, the Pierre Auger Observatory and the ANITA detector itself, both for a diffuse flux~\cite{Romero-Wolf:2018zxt,Smith:2020ecb} and for point sources~\cite{IceCube:2020gbx,ANITA:2021xxh}. Searches for these upgoing showers with the fluorescence detector of the Pierre Auger Observatory have also given negative results~\cite{PierreAuger:2023elf}.
These findings have attracted a lot of attention because, if these pulses are indeed due to upcoming air showers, they appear to be inconsistent with the Standard Model (SM), opening up explanations based on new physics beyond the SM by which showers or shower-like events develop in the atmosphere mediated by new particles such as sterile neutrinos, heavy Dark Matter or axions~\cite{Fox:2018syq,Collins:2018jpg,Heurtier:2019git,Cline:2019snp,Borah:2019ciw,Liang:2021rnv}.  

Other plausible explanations related to the mechanisms by which pulses are generated have also been suggested. One of them suggested reflections from sub-surface interfaces below the ice~\cite{Shoemaker:2019xlt}. However, no such sub-surface structures have been found in the regions where these pulses appear to come from~\cite{Smith:2020ecb}. 
The hypothesis that these pulses are due to coherent transition radiation (TR) as an energetic shower develops through the ice-air interface gave rise to two different possibilities. TR could be induced by showers starting in ice that exit into the atmosphere through the interface in the upward direction. This can be expected in the SM also from showers produced by neutrinos traveling through a large portion of an Earth's chord and producing the shower just before exiting. 
Simulations revealed again that the implied neutrino fluxes should have been observed with other instruments sensitive to UHE neutrinos such as IceCube and the Pierre Auger Observatory, and this explanation was dismissed~\cite{Motloch:2016yic}. 
Alternatively, the TR pulse could be produced by UHECR showers developing in air in a downward direction that also intercept the air-ice interface. This was postulated as a completely different possibility based on an analytical model to calculate the TR pulses induced by cosmic ray showers at moderately inclined zenith angles~\cite{deVries:2019gzs}. These results indicated that, for some particular shower geometries, the resulting pulses detected at the ANITA location could display the same polarity as direct showers, favouring a UHECR origin of the anomalous events~\cite{deVries:2019gzs}. There are, however, no detailed simulations of the coherent pulses that would develop in such situation to confirm these results.

In this article, we present for the first time a simulation of coherent transition radiation from air showers that intercept the ice surface and predict the polarity of the detected pulses at high altitudes. The simulation is based on the ZHS code~\cite{zas1992electromagnetic}. It follows the same approach used to calculate coherent TR pulses from showers that develop in the ice and exit into the atmosphere~\cite{Motloch2015}, and builds on the recent inclusion of magnetic deflections~\cite{Ammerman-Yebra:2023rhr} which are crucial to realistically simulate the pulses induced by air showers. A brief summary of the ZHS code and the TR calculation is presented in Section \ref{sec:zhs}. In Section \ref{sec:pulses} we compare the pulses produced by showers that do not intercept the ice interface with those that do. We have considered cases in which the intercept is right at shower maximum, and others in which the intercept is past shower maximum as could be expected for the geometries compatible with the two observed anomalous events. In Section \ref{sec:TR-ANITA3} we discuss the calculation of pulses which are consistent with the case examples studied in~\cite{deVries:2019gzs}, and some others, to illustrate the polarity of the pulses. Our conclusions are presented in Section \ref{sec:conclusions}.

\section{Transition Radiation with the ZHS Monte Carlo simulation code}
\label{sec:zhs}

\subsection{The ZHS Monte Carlo code}

The ZHS program~\cite{zas1992electromagnetic} is a Monte Carlo simulation of electromagnetic showers in homogeneous media specifically designed to calculate the emission of coherent radio pulses. The code accounts for particle transport in discrete steps, which are referred to as tracks, and includes bremsstrahlung, pair production, and interactions with matter electrons responsible for the excess charge in the shower, namely, M\o ller, Bhabha, Compton scattering and electron-positron annihilation. Tracks are naturally split in the Monte Carlo by the particle interactions. Further subdivisions are made to ensure that their length is below 0.1 electromagnetic radiation lengths in the medium. Multiple elastic scattering (using  Moli\`ere's theory) and ionization losses are implemented as continuous effects at each step. For low energies the tracks can be further subdivided comparing them to the particle range. Electrons, positrons and photons are followed down to a kinetic energy threshold as low as 100 keV, accounting for particle delays due to trajectory deviations and sub-luminal velocities, which are crucial for interference effects. 
The coherent electromagnetic emission from the shower is calculated summing the contributions to the radiated pulse from all the tracks due to charged particles above the energy threshold. For this purpose, the particles are assumed to move in straight lines with constant velocity between the end points of the track. Convergence of results as the step is reduced has been checked \cite{Alvarez:1995}. The contributions of these tracks to the electric field can be calculated at a given observing position both in the frequency~\cite{Alvarez-Muniz:2010wjm} and time~\cite{zas1992electromagnetic} domains, taking into account the phases or time delays with respect to a particle moving along shower axis at the speed of light. 

The calculation of the radio pulse relies on the ZHS algorithm~\cite{zas1992electromagnetic}. 
Without loss of generality, the shower is assumed to develop in a direction making an angle $\psi$ relative to the $z$ axis perpendicular to the interface, starting at $z=0$ and with the interface at $z=z_{tr}$,(see Fig.~\ref{fig:Geometry}). 
The radio signal is obtained summing the contributions to the vector potential due to the particle tracks in the shower. The contribution of a single particle track moving with constant velocity $v$, such that $\beta=v/c$ with $c$ the light speed in vacuum, in a medium of refractive index $n$ is given by~\cite{Alvarez-Muniz:2010wjm}:
\begin{equation}
\textbf{A} = \frac{\mu e}{4\pi R}
\,\textbf{v}_\perp\, 
\frac{\Theta\left[t-\frac{nR}{c}-(1-n\beta\cos\theta)t_1\right]-\Theta \left[t-\frac{nR}{c}-(1-n\beta\cos\theta)t_2)\right]}{(1-n\beta\cos\theta)}\,\,\,  .
\label{eq:zhs_time_vp}
\end{equation}
Here, $t$ is the time measured by the observer, $t_1$ and $t_2$ are the start and end times of the particle track, $\theta$ is the angle between the position vector of the observer relative to the center point of the track, $\textbf{R}$, and the particle velocity, and $R$ (the modulus of $\textbf{R}$) is the corresponding distance. The arguments of the two Heaviside functions, $\Theta$, relate the retarded times at the observer with the times of the start and end points of the track. The retarded time from a given track is obtained in the Fraunhofer limit, but by allowing $\textbf{R}$ to be different for each track, it is possible to calculate the coherent pulses also when the observer is not in the Fraunhofer limit with respect to the shower \cite{Garcia-Fernandez:2012urf}. All that is needed is that the Fraunhofer approximation is valid for all tracks, and this is achieved by reducing the size of the tracks in the calculation \cite{Garcia-Fernandez:2012urf}. 
The electric field is obtained taking minus the time derivative of the final vector potential. 

\subsection{Transition Radiation with the ZHS code}

The original ZHS program was initially designed for ice, ignoring particle deflections in the geomagnetic field and to work in the frequency domain only. It was later extended to deal with other media~\cite{AlvarezMuniz2009}, including air, and to calculate pulses in the time domain~\cite{AlvarezMuniz2012}. 
A special version of the ZHS program was developed in~\cite{Motloch2015} 
to account for the coherent TR generated as a shower that starts in ice, traverses the ice-air interface, and continues developing in the atmosphere~\cite{Motloch2015}. 
This program is capable of calculating coherent TR for an observer that is located in the air (after the ice-air interface), and it has been shown that the neutrino energy necessary for such pulses to be compatible with the anomalous events detected with ANITA, would have to be even larger than in the case of tau leptons decaying in flight, ruling out TR induced by showers developing from ice to air as a possible explanation~\cite{Motloch:2016yic}. 
However, this version of ZHS cannot be used to simulate realistic pulses from showers developing in air and penetrating the ice sheet for an observer in the atmosphere, in order to test the postulated explanation of the anomalous events based on TR in~\cite{deVries:2019gzs}. 
A realistic simulation requires that magnetic deflections of charged particle tracks are accounted for, as recently implemented and described in~\cite{Ammerman-Yebra:2023rhr}. 
This program has been used to study properties of pulses in air under different conditions of density and magnetic field intensity.

In this work, we describe a special version of ZHS building on the ZHS with magnetic deflections~\cite{Ammerman-Yebra:2023rhr} to simulate the TR pulse generated when an air shower intercepts the ice surface, assumed flat, with the electric field observed in air at high altitude. 
The program follows closely the method described in~\cite{Motloch2015}, summing both the direct and the reflected components emitted from the tracks in air that contribute at the observer's position, and the refracted component emitted from the tracks inside the ice towards the observer. 
The program is limited to electromagnetic showers and to constant density air. 
However, by choosing a value of the density which is similar to air density at the ice-air interface, corresponding to about 3 km altitude at the South Pole, it is possible to simulate showers that have a distribution of particles that approximately reproduces that of showers induced by cosmic rays in the atmosphere. 
The details of the simulation cannot be expected to fully match those of a realistic cosmic ray shower developing in a more realistic atmosphere with density decreasing exponentially with altitude. However, it has been shown that many aspects of pulses from air showers are well reproduced with the ZHS which allows quantitative predictions at the level of $\sim 20~\%$~\cite{Ammerman-Yebra:2023rhr}. It is therefore anticipated that these limitations will not impact the polarity of the pulses.

The procedure to calculate transition radiation has been described in \cite{Motloch2015}. 
The shower is developed in two stages. Firstly, the shower is simulated in air and when a particle track crosses the ice-air interface it is split in two pieces with the first part ending at the interface. 
The time, position, velocity, direction and particle type at the interface are stored for further processing. In this way, charged particle tracks are classified in two sets, those that fully develop in the first medium and those that fully develop in the second one. 
For the simulation in ice the particles are taken from the recorded buffer and injected at the interface position, using the recorded parameters, velocity, energy and particle type to resume shower development. 

\begin{figure}[ht]
	\centering
    	\subfigure{\includegraphics[width=0.49\linewidth]{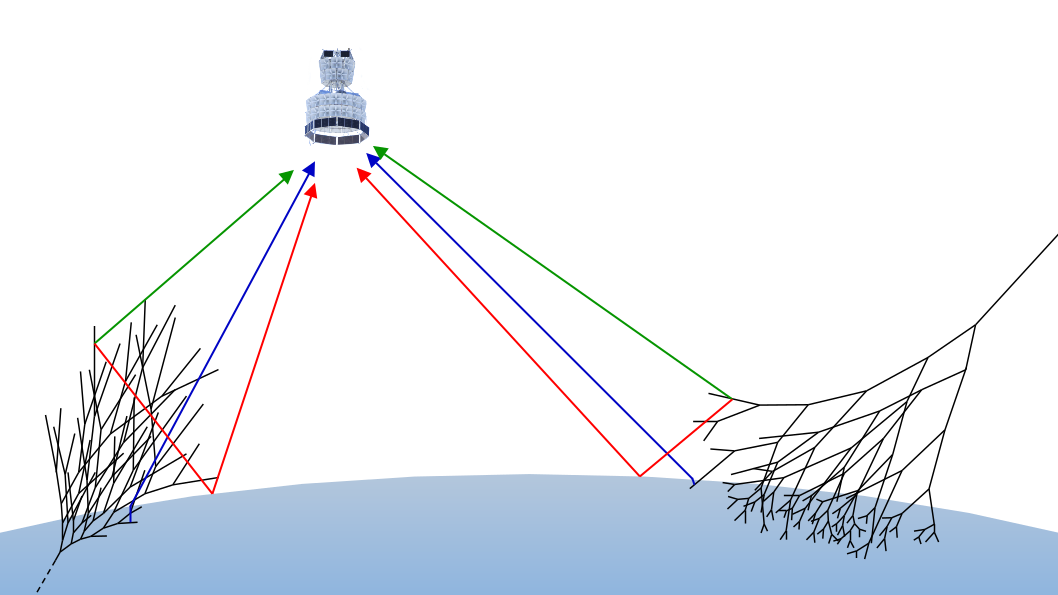}}
    	\subfigure{\includegraphics[width=0.49\linewidth]{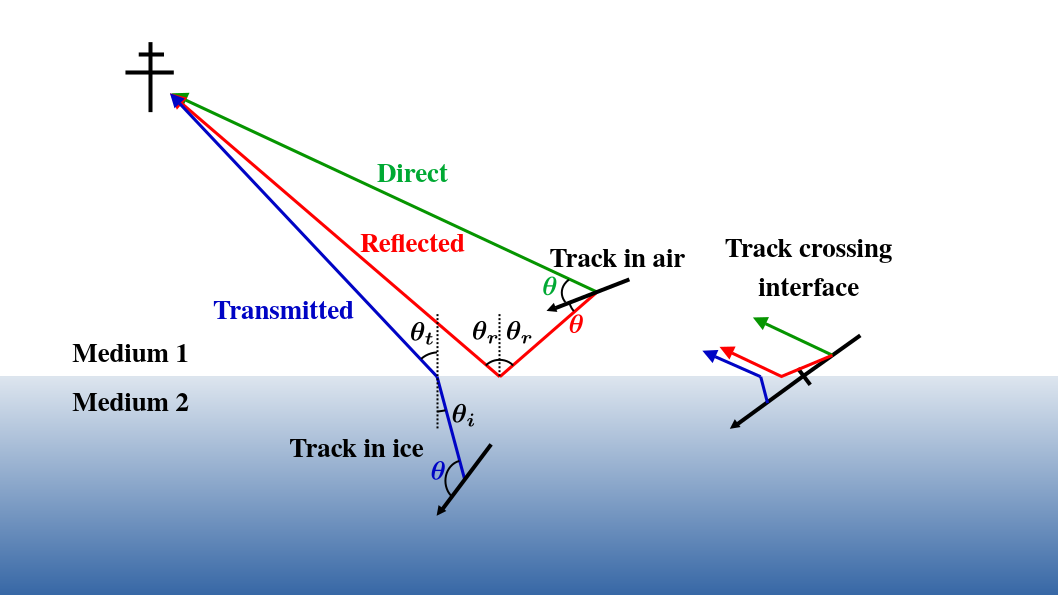}}
\caption{Sketch of possible scenarios for an event inducing transition radiation (TR) detected at the altitude of the ANITA detector (not to scale). The three contributions to the radio pulse are indicated with three different colors. For the tracks in air the  direct (green) and reflected (red) contributions are relevant and for the tracks in ice only the transmitted (refracted) contribution is needed (in blue). The relevant emission angle  $\theta$ needed to calculate the tack contribution from Eq.~(\ref{eq:zhs_time_vp}) is shown for all cases. In addition, the incidence angle at the interface is needed for both the reflected and transmitted contributions to calculate the distance $R$ in Eq.~(\ref{eq:zhs_time_vp}) and the corresponding Fresnel coefficient (see text).}
\label{fig:sketch_contributions}
\end{figure}
%
The contributions to the electric field from each shower track can be classified into direct, reflected or refracted, depending on the medium where the track develops and the location of the observer, as shown in 
 Fig.\,\ref{fig:sketch_contributions}. 
For a detector in the atmosphere (the ANITA case) the emission from tracks in the air can reach the detector both directly or after reflection at the ice boundary, while for the tracks in ice only the refracted component at the boundary reaches the detector as sketched in Fig.\,\ref{fig:sketch_contributions}. 
For each track, the amplitudes of the refracted or the reflected components must be calculated after multiplying the incident radiation at the interface by the corresponding Fresnel coefficients. 
Care has to be taken to account for the correct time delays or phase shifts of the refracted and reflected components. The developed method ensures that the boundary conditions at the dielectric interface are satisfied~\cite{ginzburg1982transition,James:2010vm}.
For a shower transiting from air to ice, the tracks in the first medium contribute with both the direct and reflected components, while for those in the second medium only the component refracted at the boundary reaches the detector.    

The calculation of TR, requires choosing the relevant emission angle and distance from the track to the detector in accordance with whether we are dealing with the direct, reflected, or refracted contributions, as shown in Fig.~(\ref{fig:sketch_contributions}). The direct emission from tracks in air uses directly Eq.~(\ref{eq:zhs_time_vp}), where the angle $\theta$ is that between the track direction and the ray that goes in a straight path from the middle of the track to the observer. The corresponding angles and distances when calculating the reflected and refracted components can be easily obtained with the help of ray tracing. Each track contribution is based on Eq.~(\ref{eq:zhs_time_vp}) but the distance $R$ must account for the extra path traveled by the ray that departs from the middle of the track and reaches the observer after reflection or refraction. The angle $\theta$ is that between the emitted ray and the direction of the track, and the emission direction must be such that the reflected or refracted ray passes by the detector position. The angle $\theta$ for the refracted contribution must be obtained numerically unless the observer is in the Fraunhofer limit. As stated before, the electric field amplitude must be multiplied by the Fresnel factor taking into account the corresponding angle of incidence at the interface (see Fig.~\ref{fig:sketch_contributions}).

This calculation does not allow us the separation of the emission due to TR from that due to the shower. One could try to separate the contributions from tracks that reach the interface from those that are fully contained in air. This procedure however introduces spurious contributions from the artificial (and arbitrary) separation of the tracks reaching the interface. Large cancellations can be expected when combining the tracks that reach the interface with those that develop fully in the air, so no identification of the TR contribution itself is possible. To quantify the effect of TR we will instead compare showers in which the expected amount of TR increases by placing the interface at three different depths along the shower development corresponding to a shower that is fully contained in air (``no-TR''), a shower intercepting ice at shower maximum (``max-TR''), and a shower crossing the interface well after reaching shower maximum, when shower size is close to half its maximal value (``mid-TR''). 

\section{Radio pulses from air showers intercepting the surface of the ice}
\label{sec:pulses}

\begin{figure}[ht]
\centering
\includegraphics[width=0.85\linewidth,trim={2cm 2cm 2cm 2cm},clip]{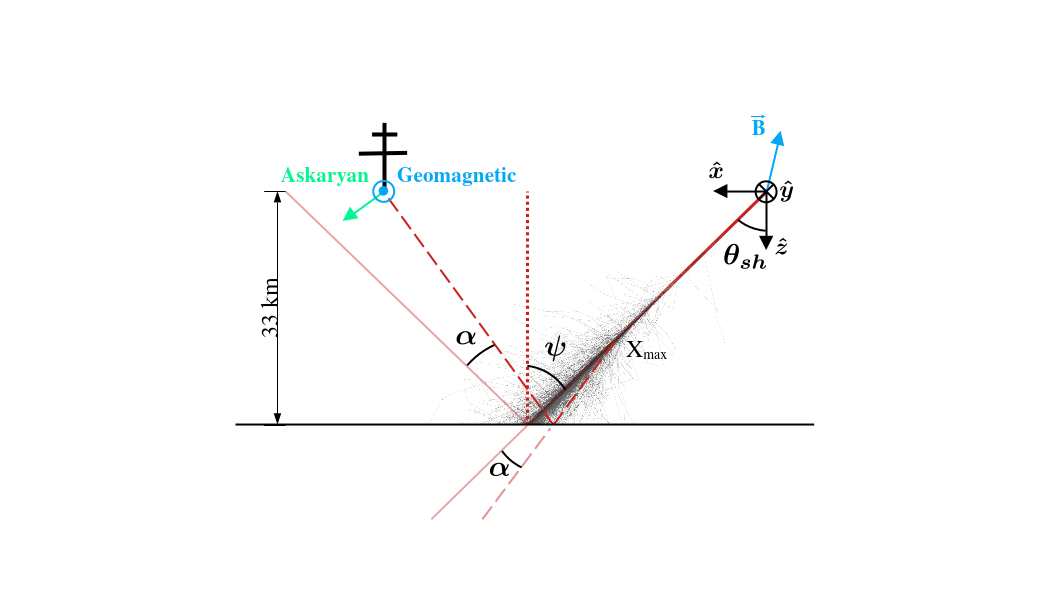}
\caption{Sketch of the chosen set-up for the simulations. The observer at an altitude of 33 km, and the magnetic field are chosen to be in the $\boldsymbol{\hat{x}}\boldsymbol{\hat{z}}$ plane which is the plane of this figure. The transverse current and the geomagnetic component of the electric field are orthogonal to this plane. The zenith angle of the shower $\psi$ is depicted, while $\alpha$ is the off-axis angle of the emission reaching the observer (relative to the shower axis) defined with respect to the position of shower maximum.}
\label{fig:Geometry}
\end{figure}

To illustrate the effects of TR we have fixed the geometry of the shower relative to the air-ice interface to match the example studied by de Vries et al.~\cite{deVries:2019gzs} as sketched in Fig.\,\ref{fig:Geometry}. 
The coordinate system has the $z$-axis normal to the interface and points to it from the first medium. The shower axis lies in the $\boldsymbol{\hat{x}}\boldsymbol{\hat{z}}$ plane, and the shower zenith angle $\theta_\text{sh}=\psi=55^\circ$ also shown in Fig.\,\ref{fig:Geometry}, was chosen to match the example in \cite{deVries:2019gzs}.
The magnetic field, also contained in the $\boldsymbol{\hat{x}}\boldsymbol{\hat{z}}$ plane, is chosen to have a strength of 0.65~G with an inclination of $71^\circ$ relative to the $x$-axis. These are typical values at Antarctica. We place the observer in air, in the $\boldsymbol{\hat{x}}\boldsymbol{\hat{z}}$ plane, at an altitude of 33 km above the ice surface. Its exact position is calculated so that it is intercepted by the reflection of a ray that is emitted at the position of shower maximum making an angle $\alpha$ relative to shower axis (the \textit{off-axis} angle) as indicated in Fig.\,\ref{fig:Geometry}. When the off-axis angle is close to the Cherenkov angle, $\alpha_C$, the coherence of the reflected emission, including the component of TR at the interface, is expected to be maximal. 
For this particular geometry the transverse current due to the Lorentz force, $\boldsymbol{J_\perp}$, is perpendicular to the $\boldsymbol{\hat{x}}\boldsymbol{\hat{z}}$ plane, pointing into the negative $\boldsymbol{\hat{y}}$ direction and therefore the $E_y$ component of the field is dominated by the geomagnetic effect.
The orthogonal polarization component, $E_{\tilde{x}}$, lies in the $\boldsymbol{\hat{x}}\boldsymbol{\hat{z}}$ plane, in a direction perpendicular to the emitted ray which depends on its off-axis angle, with unit vector $\tilde{x}$, and is typically smaller than $E_y$ by a large factor as it is due to the excess charge in the shower or the Askaryan effect. By choosing this geometry the geomagnetic effect and the excess charge or Askaryan effect are separated in these two orthogonal polarizations. The relative position of the plane containing the shower, the ice surface and the magnetic field will be kept constant for all the simulations that will be described in the following sections.

For air we have used a density $0.9095 \cdot 10^{-3}\,\text{g\,cm}^{-3}$ corresponding to the approximate density of the atmosphere at 3 km of altitude above sea level, a typical altitude of the ice surface at Antarctica, with a refractive index of 1.0002255 and a corresponding Cherenkov angle of $1.22^\circ$. For ice we choose a density of $0.42\,\text{g\,cm}^{-3}$ and a refractive index of 1.36 (Cherenkov angle $\alpha_C=42.7^\circ$) consistent with the properties of the firn layer at the South Pole~\cite{kravchenko2004situ}. 
We have calculated pulses induced by 1 PeV electron showers in the geometry described in Fig.\,\ref{fig:Geometry} for different off-axis angles $\alpha$. By choosing the starting point of the showers at different positions relative to the interface, we have considered air showers that fully develop in air before reaching the ice and do not produce TR (no-TR), and showers that intercept the air-ice interface at two different development depths, at $750~\text{g\,cm}^{-2}$ when the shower size drops to about half its maximum value as an intermediate case (mid-TR) and at $520\,\text{g\,cm}^{-2}$, i.e. close to shower maximum for a 1 PeV electron shower, for expected maximal TR (max-TR).

\begin{figure}[ht]
	\centering
    \subfigure[Geomagnetic component ($E_y$)]{\includegraphics[width=0.49\linewidth]{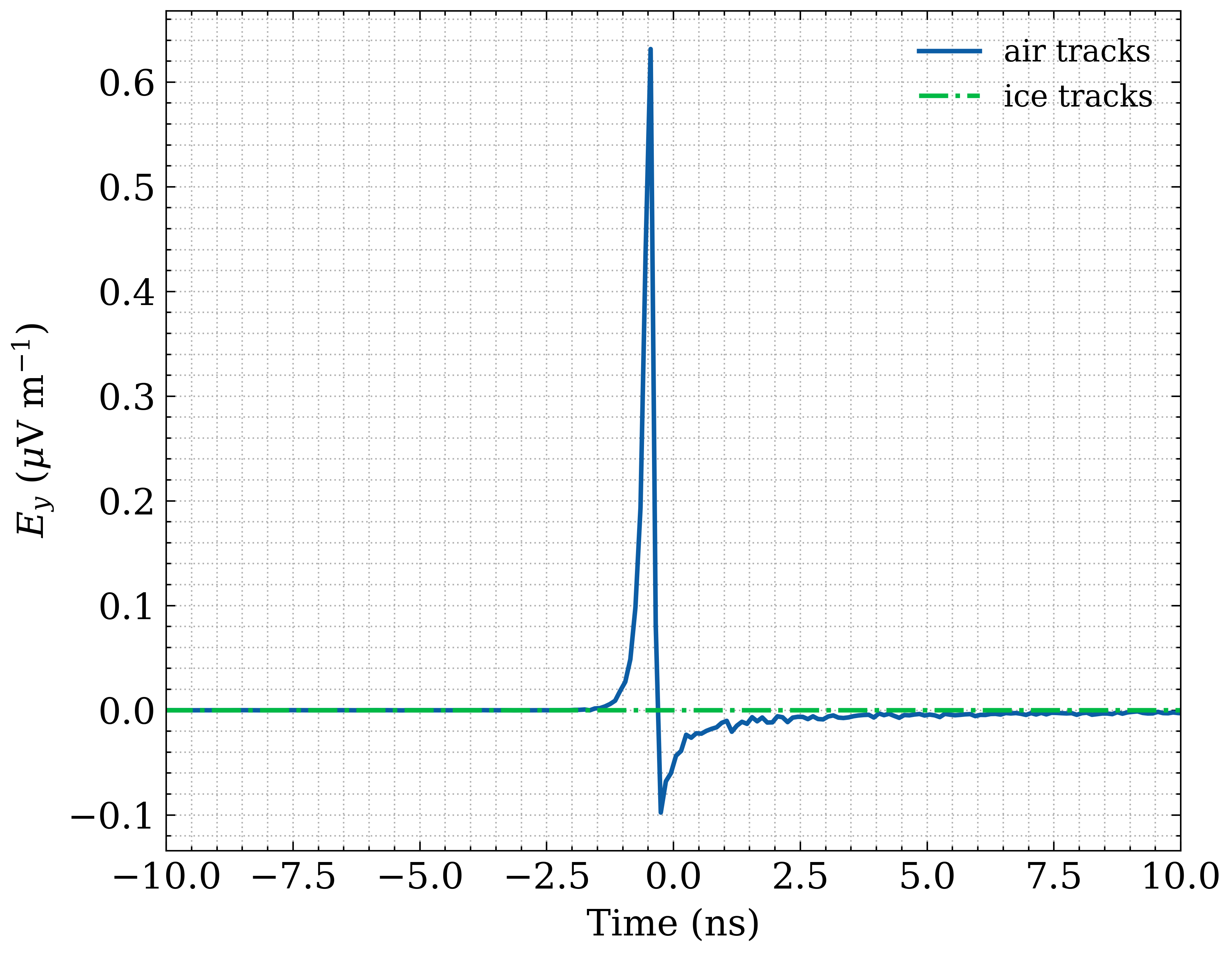}}
    \subfigure[Askaryan component ($E_{\tilde{x}}$)]{\includegraphics[width=0.49\linewidth]{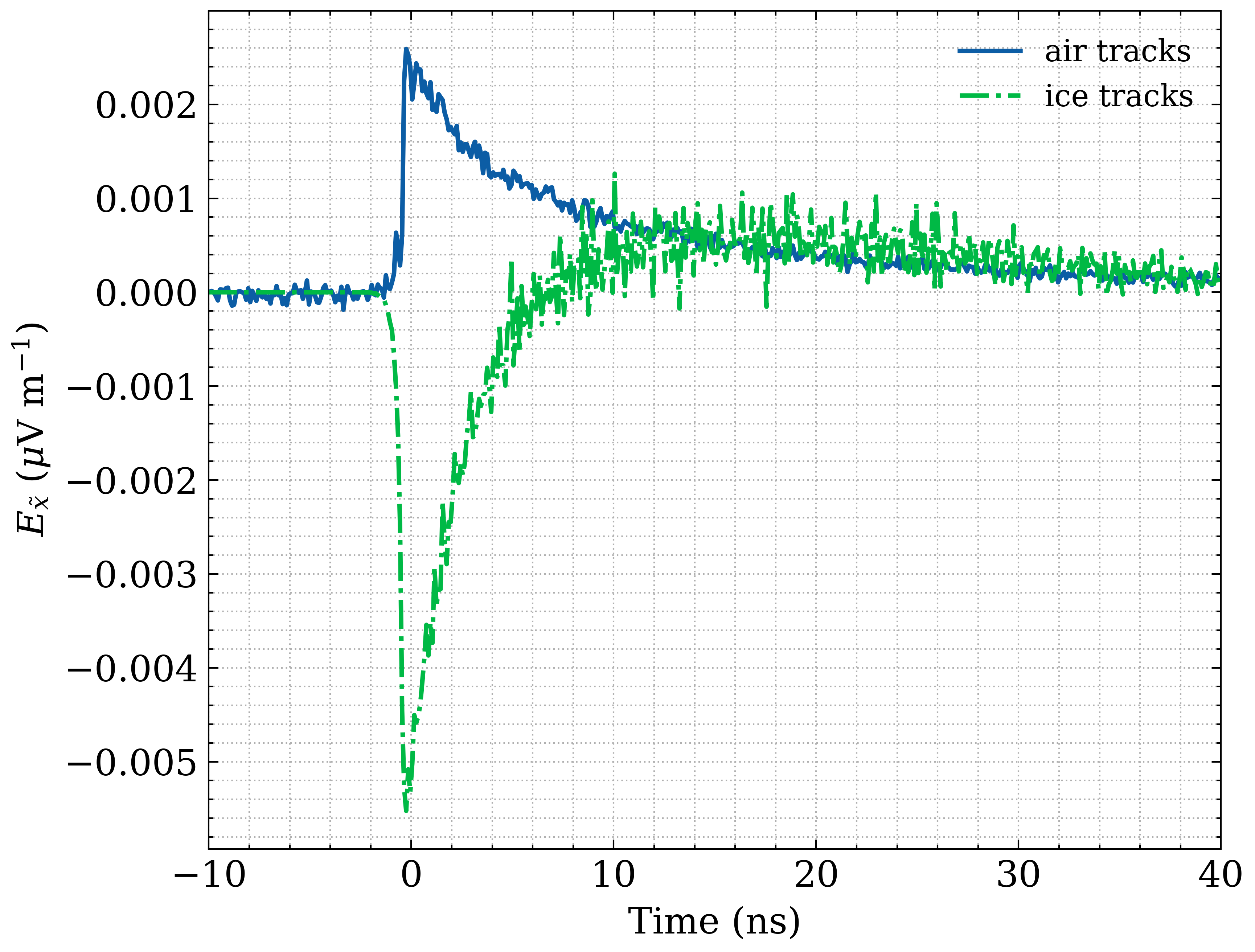}}
    \caption{Comparison of the contributions to the electric field of the $y$ (geomagnetic, left panel) and $\tilde{x}$ (mostly due to Askaryan, right panel) polarizations,  induced in a 1 PeV electron-induced air shower simulated with ZHS, that crosses the ice-air interface at shower maximum (case ``no-TR'', see text). The contributions from the tracks in air and in ice are shown separately in each panel. The observer has been placed at an off-axis angle corresponding to the Cherenkov angle ($\alpha=\alpha_C\sim 1.22^\circ$, see Fig.\,\ref{fig:Geometry}).}
\label{fig:CherenkovPulses}
\end{figure}

The results of our simulations are displayed in Fig.~\ref{fig:CherenkovPulses} for a shower that intercepts the surface at shower maximum (max-TR) and for an observer in the Cherenkov direction ($\alpha=\alpha_C\sim 1.22^\circ$) as shown in Fig.\,\ref{fig:Geometry}. 
In Fig.\,\ref{fig:CherenkovPulses}a, we display the component of the field parallel to the $y$-axis, $E_y$, which, adopting the geometry in Fig.\,\ref{fig:Geometry}, is dominated by the geomagnetic effect. In Fig.\,\ref{fig:CherenkovPulses}b, we display the component $E_{\tilde{x}}$, mostly due to the Askaryan effect.
For each of these two components, $E_y$ and $E_{\tilde{x}}$, we separate the contributions due to the tracks in the air, shown in blue (corresponding to the direct and reflected contributions), from those tracks fully in ice plotted in green (refracted contribution), see Fig.\,\ref{fig:sketch_contributions}. 

As shown in Fig.\,\ref{fig:CherenkovPulses}, the geomagnetic effect dominates in air when compared to ice because the density is $\sim 1000$ times lower and the transverse currents are much larger, in fact, $E_y$ is close to two orders of magnitude larger than $E_{\tilde{x}}$. 
The relative contribution to $E_y$ from the tracks in ice displayed in Fig.\,\ref{fig:CherenkovPulses}a is shown to be negligible, and the bulk of coherent TR can be obtained keeping only the contributions of the tracks in air~\cite{revenu2012radio, garcia2018calculations, garcia2019influence}.
However, the tracks in ice contribute mostly to the $E_{\tilde{x}}$ polarization (after refraction) as shown in Fig.\,\ref{fig:CherenkovPulses}b, because the Askaryan effect dominates in ice. This component also displays a much wider time spread of 10's of ns compared to the narrow few ns of $E_y$. This is because in ice the Cherenkov angle where the emission peaks corresponds to an off-axis angle of about $\alpha_C=43^\circ$, and radiation with such angle cannot exit into the air because it either propagates downward due to the gradient of refractive index of the firn, or it undergoes total internal reflection at the interface~\cite{deVries:2019gzs}.
As a consequence, the emission in ice is only transmitted for off-axis angles well away from the Cherenkov angle where the pulses are much broader in time~\cite{Alvarez-Muniz:2010wjm}. 
Although not explicitly displayed in Fig.\,\ref{fig:CherenkovPulses}, the direct emission to the observer is very small relative to the reflected component because it corresponds to an off-axis angle much greater than the Cherenkov angle in air. As a result, it is also possible to approximate the bulk of coherent TR in this geometry by considering only that coming from the tracks in air and only accounting for the reflected contribution~\cite{deVries:2019gzs}. 

\begin{figure}[ht]
	\centering
	\centering
    	\subfigure[$\alpha=\alpha_C=1.22^\circ$]
     {\includegraphics[width=0.31\linewidth]{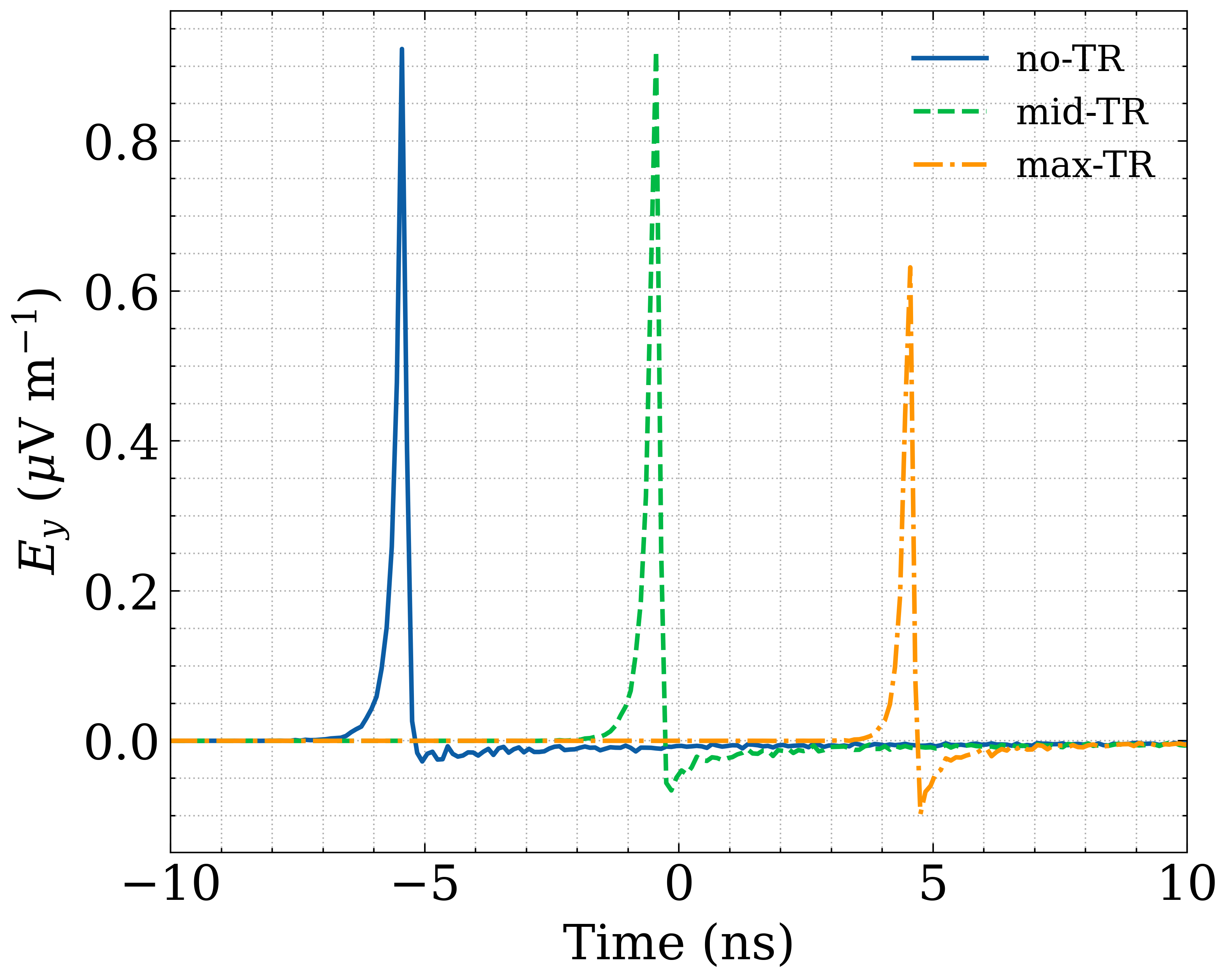}}
        \subfigure[$\alpha=\alpha_C-0.2^\circ=1.02^\circ$]
        {\includegraphics[width=0.33\linewidth]{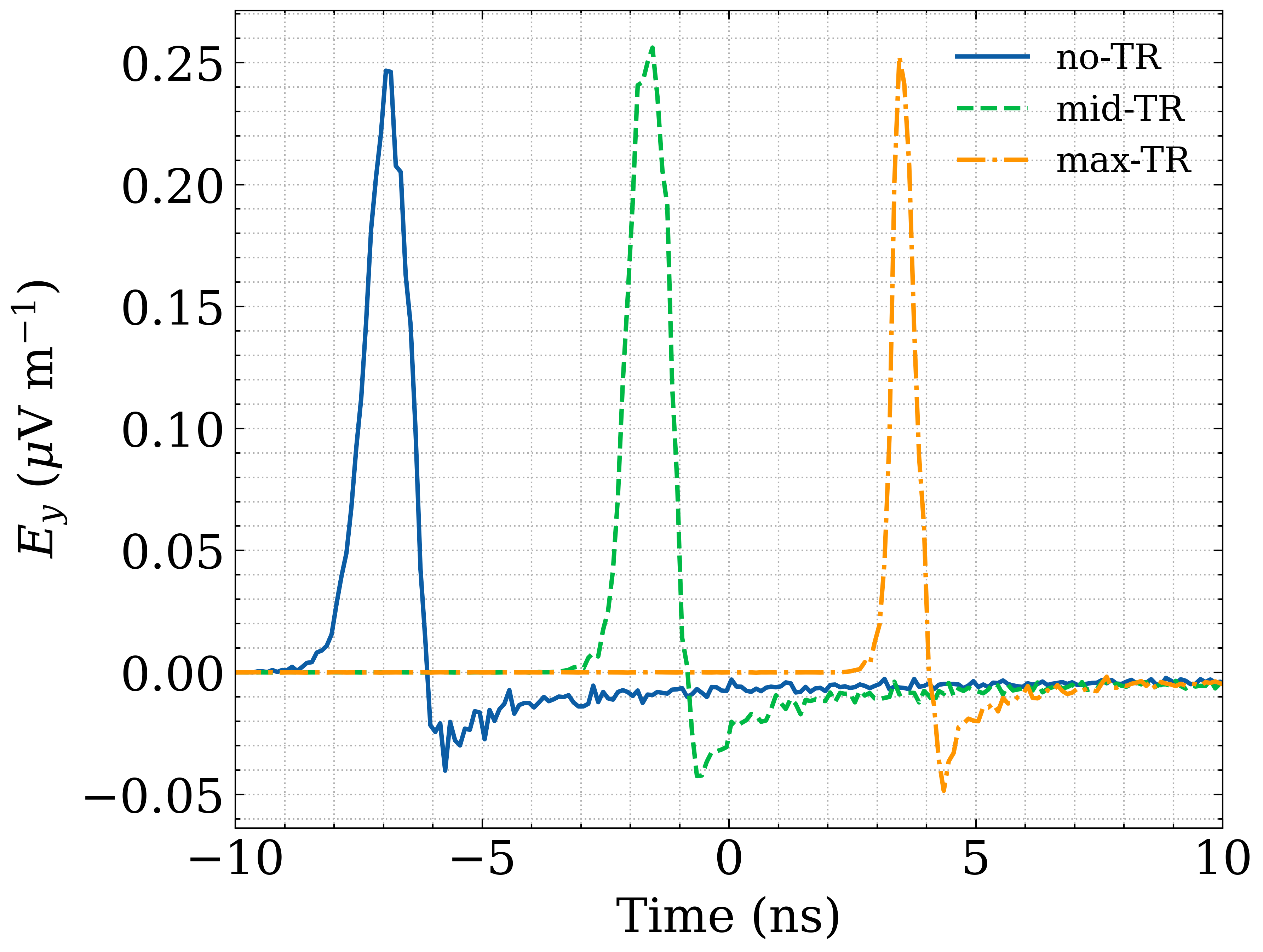}}
        \subfigure[$\alpha=\alpha_C+0.2^\circ=1.42^\circ$]
        {\includegraphics[width=0.32\linewidth]{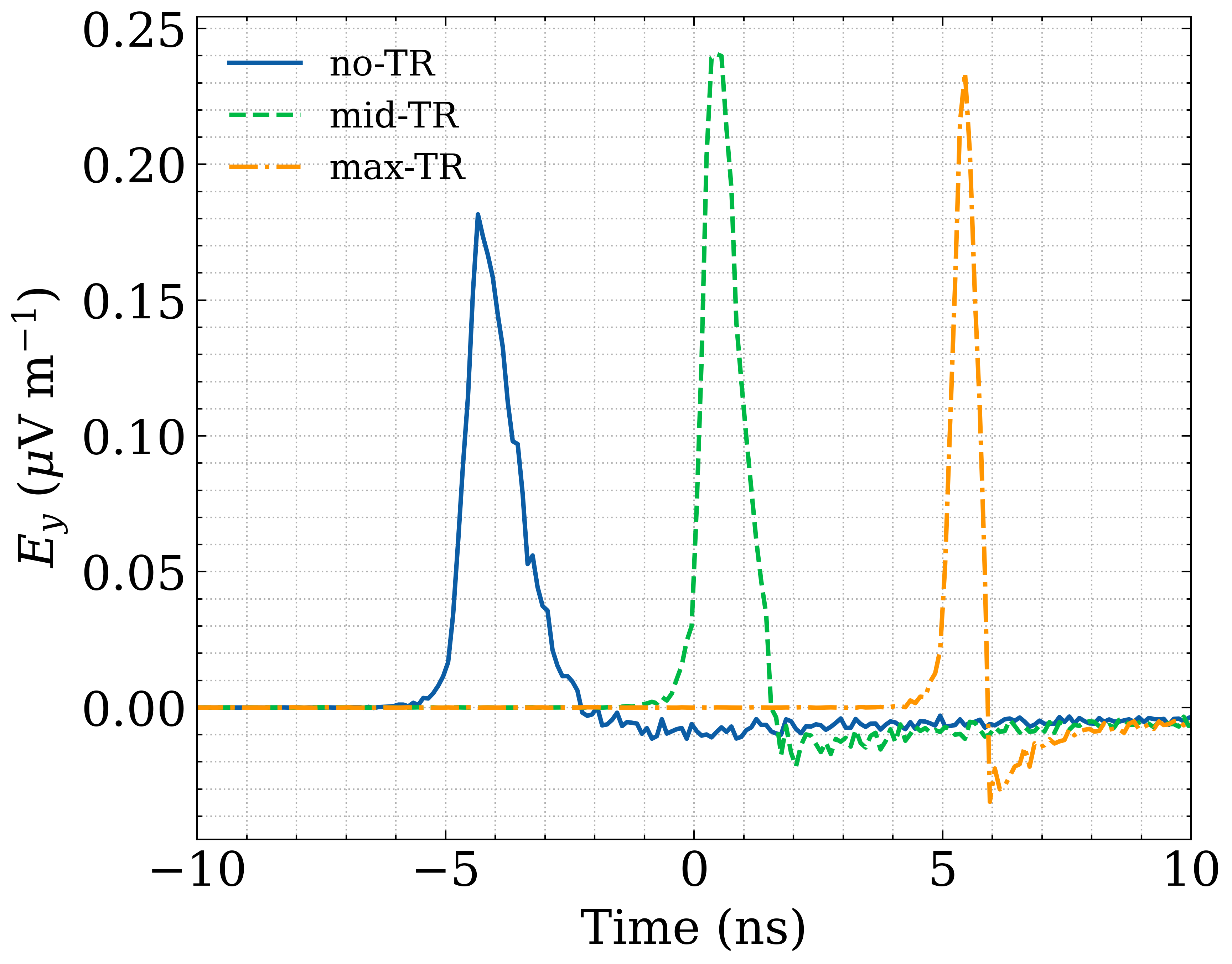}}    
\caption{Time traces of the $E_y$ polarization of radio pulses for three 1 PeV air showers simulated with ZHS for different geometries: A shower that does not intersect the air-ice interface (no-TR), a shower that intersects the interface after $X_{\rm max}$ such that only roughly half of the maximum number of particles in the shower cross the interface (mid-TR), and a shower that crosses at $X_{max}$ (max-TR). The time traces of cases no-TR and max-TR have been arbitrarily shifted by $\pm 5$~ns for better visualization. Each of the three panels (a), (b) and (c) compares the emission for observers located so that the off-axis angle of the emission (measured at shower maximum, see Fig.\,\ref{fig:Geometry}) are respectively $\alpha=1.22^\circ$  (Cherenkov angle in air), $1.02^\circ$ and $1.42^\circ$, i.e. $0.2^\circ$ ``inside'' and ``outside'' the Cherenkov cone respectively.}
\label{fig:time_tr_comparison}
\end{figure}
\begin{figure}[ht]
	\centering
    	\subfigure[$\alpha=\alpha_C=1.22^\circ$]
     {\includegraphics[width=0.32\linewidth]{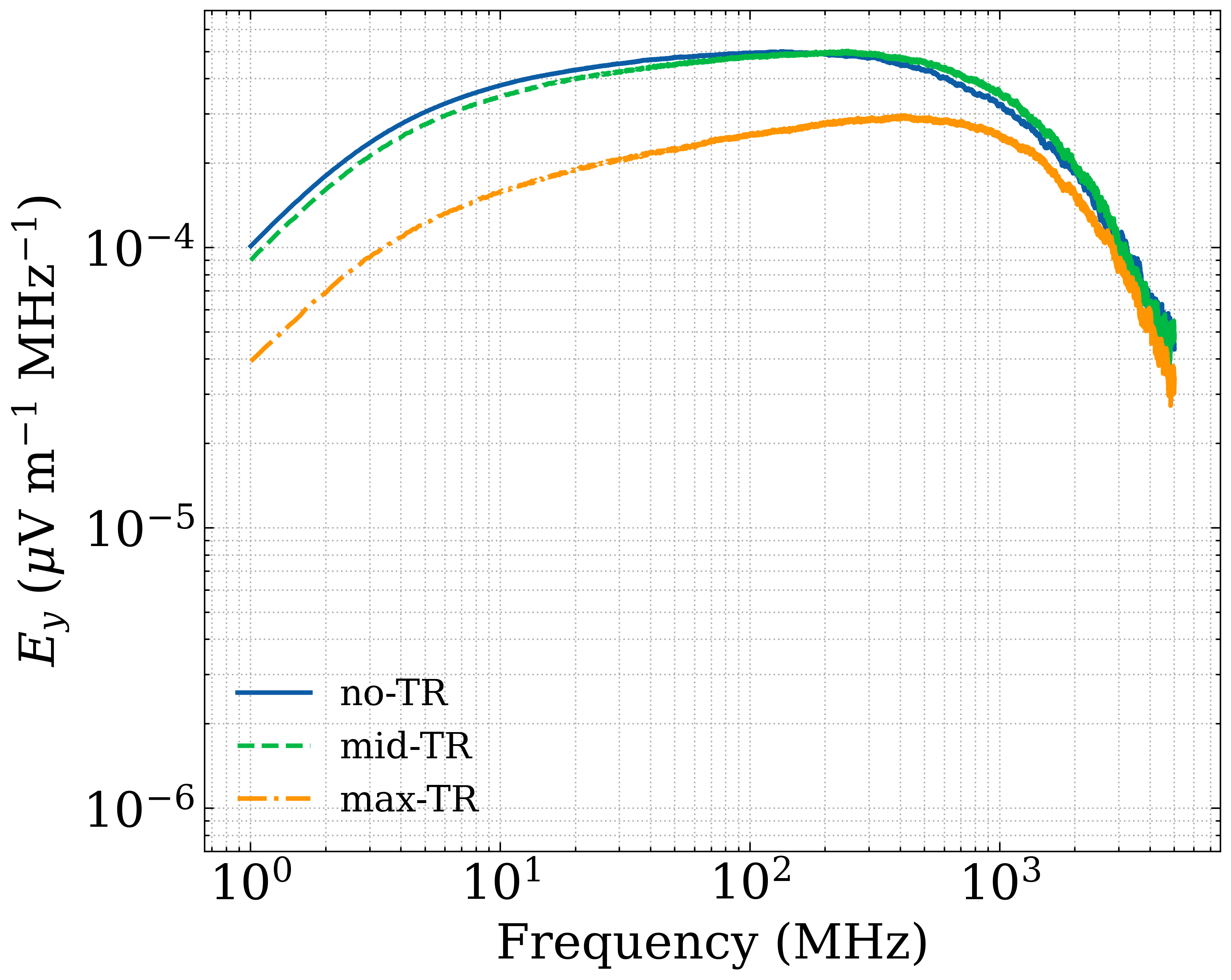}}
        \subfigure[$\alpha=\alpha_C-0.2^\circ=1.02^\circ$]
        {\includegraphics[width=0.32\linewidth]{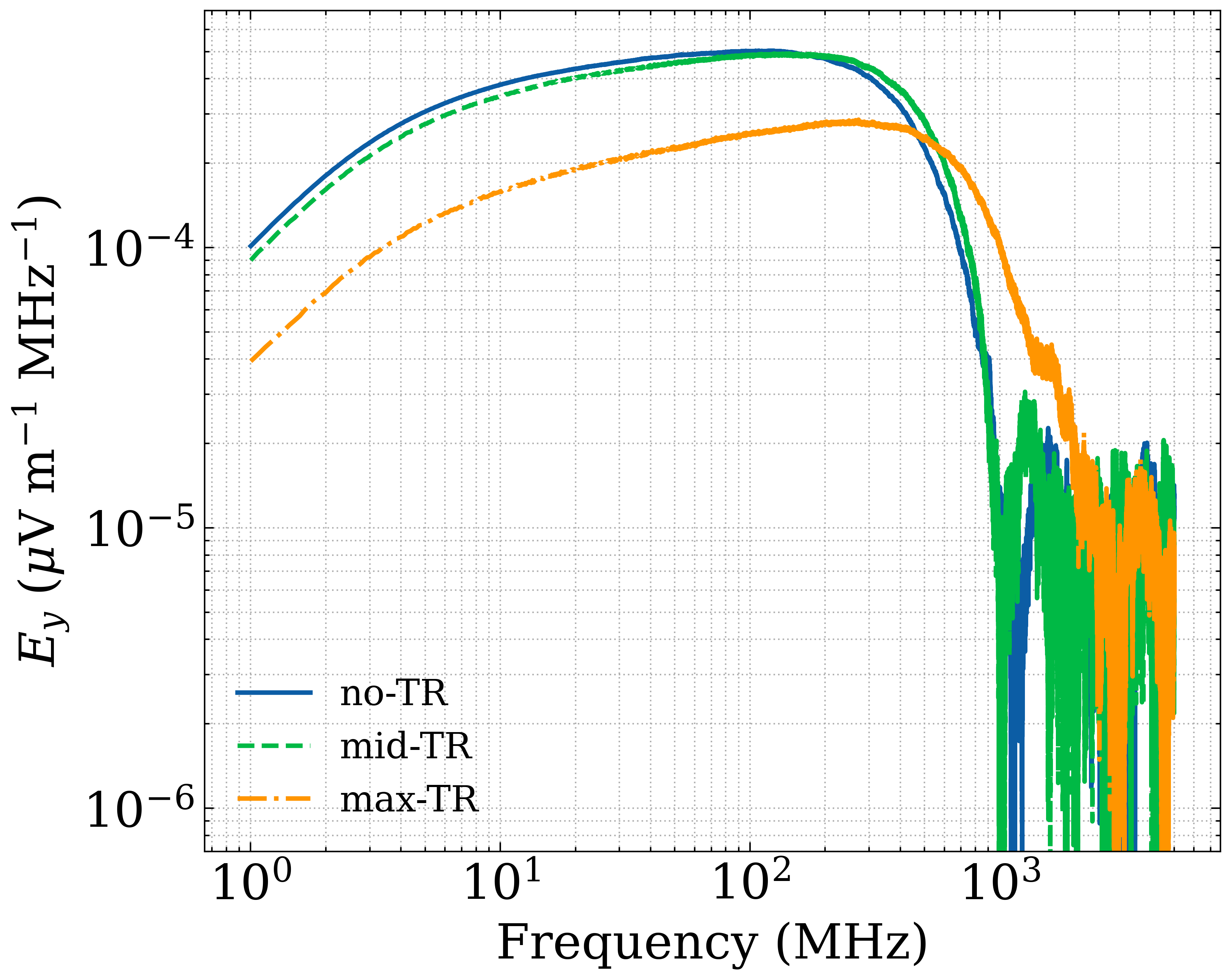}}
        \subfigure[$\alpha=\alpha_C+0.2^\circ=1.42^\circ$]
        {\includegraphics[width=0.32\linewidth]{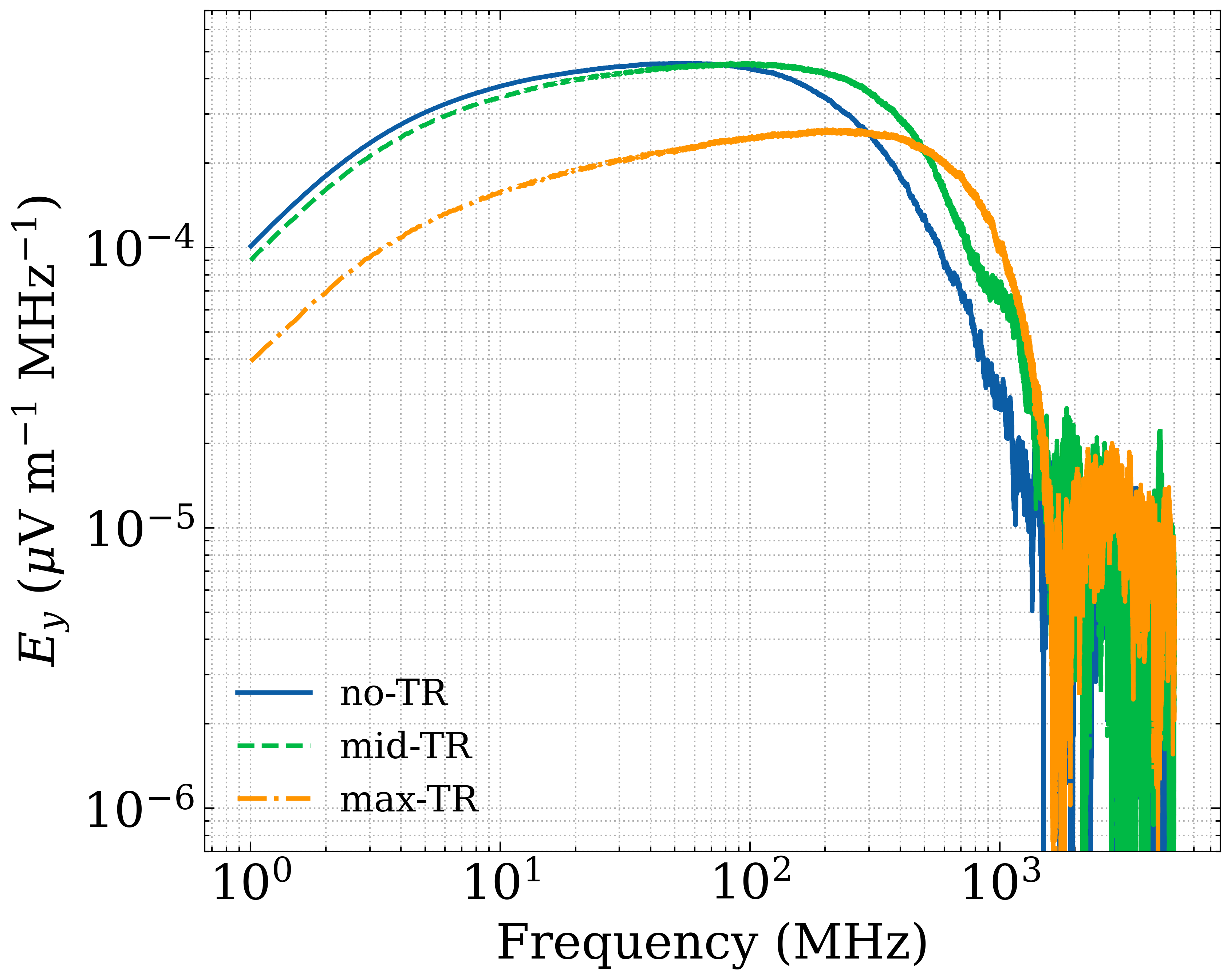}}
\caption{Frequency spectra of the $E_y$ polarization of a radio pulse for the same cases as in Fig.\,\ref{fig:time_tr_comparison}.}
\label{fig:freq_tr_comparison}
\end{figure}

To further visualize the effects of TR we compare in Fig.\,\ref{fig:time_tr_comparison} the pulses received at a high-altitude observer for three shower geometries, no-TR, mid-TR and max-TR, and for different off-axis angles. Only the dominant $E_y$ polarization is shown.
In panel (a) the observer is located so that the pulse is emitted and reflected with an off-axis angle $\alpha=\alpha_C=1.22^\circ$ equal to the Cherenkov angle in air; in panel (b) $\alpha=1.02^\circ$ (``inside'' the Cherenkov cone); and in panel (c) $\alpha=1.42^\circ$ (``outside'' the Cherenkov cone). 
The first remarkable observation in Fig.\,\ref{fig:time_tr_comparison} is that the radio pulses are quite similar in shape for all the three shower geometries and for all observing off-axis angles. They are all bipolar and with the same polarity, first positive and then negative, and with the second negative peak typically less intense and wider in time. 

In the case of no-TR, when the shower is fully contained in the atmosphere and the bulk of radiation reaches the observer after reflection, the pulse exhibits the same polarity as the pulses for the showers with TR (mid-TR and max-TR). According to our simulations, the polarity in any of the cases with TR is similar to that of a reflected pulse. It does not match that of the pulse produced by a particle moving in the upward direction and showering in the atmosphere after crossing the Earth, which arrives directly without reflection. This clearly disfavors the interpretation of the anomalous ANITA events as being due to UHECR showers moving in the downward direction and crossing the air-ice interface \cite{deVries:2019gzs}.

The peak amplitudes of the pulses in the case of max-TR are less than half the value obtained for the case of no-TR for all observation angles. This is not surprising because only the first part of the shower develops in air so that the total tracklength of the charged particles is reduced relative to the case no-TR\,\footnote{The amplitude of the coherent pulses is known to be proportional to the tracklength~\cite{zas1992electromagnetic}}. In all the three panels, it can also be observed that the amplitude of the second pulse (plotted in the negative direction) is smallest for the case of no-TR and largest for the case of max-TR. This is consistent with TR increasing with the number of particles crossing the interface as anticipated~\cite{Motloch2015}. 

As the shower intercepts the ice surface there is an  abrupt change of the vector potential which results in pulses that are narrower in time and display more abrupt slopes relative to showers that have no TR. 
The reduction in width in the case of mid-TR relative to that of no-TR is rather small, while for max-TR it is more significant, as can also be appreciated in the time-domain traces in Fig.\,\ref{fig:time_tr_comparison}. This is easier to recognize for emission with an off-axis angle different from the Cherenkov angle because the pulse width reflects the longitudinal shower development \cite{Alvarez-Muniz:2020ary} which is truncated when there is TR. However, the lateral spread and the time delays of the shower dominate over the longitudinal delays in the Cherenkov direction and close to it~\cite{zas1992electromagnetic,Alvarez-Muniz:2015ayz,Alvarez-Muniz:2020ary}, making these effects less evident. This effect is best seen in the frequency domain. 
In the three panels of Fig.\,\ref{fig:freq_tr_comparison} we display the corresponding frequency spectra for the same cases no-TR, mid-TR and max-TR, and off-axis angles as in Fig.\,\ref{fig:time_tr_comparison}.   
They clearly illustrate that the pulses contain more high-frequency components as the number of particles intercepting the interface increases, from cases with no TR to those with maximal TR.

\section{Transition radiation as an explanation of the anomalous ANITA III event}
\label{sec:TR-ANITA3}

The explanation of the anomalous ANITA events based on TR from UHECR showers intercepting the air-ice interface which is given in~\cite{deVries:2019gzs}, relies on the results of an analytical calculation based on a 3-dimensional model of the shower current which is abruptly cutoff with a Heaviside function as the shower intercepts the air-ice interface. The resulting pulses are calculated as the superposition of two terms, one due to the derivative of the vector potential induced by the current associated to shower development, and the other due to the derivative of the Heaviside function describing the abrupt end of the shower at the interface. The two terms correspond to pulses that can be approximated as being monopolar with opposite polarities. It is argued that an inversion of polarity is obtained from the sum of these two pulses that are not fully synchronized in time, with the pulse associated to TR taking place earlier or later with respect to the shower pulse depending on whether the detected field corresponds to emission at off-axis angles smaller or larger than the Cherenkov angle, respectively~\cite{deVries:2019gzs}. 

In this Section we apply our simulation to the particular case of the anomalous ANITA III event, adopting the same geometry as in the detected event and studying in detail its properties. A pulse detected with the ANITA instrument must have high-frequency components to trigger and therefore, it must be emitted (before reflection) with an off-axis angle close the Cherenkov angle in air. To be compatible with the anomalous event detected in the ANITA III flight it must be received with elevation angle of $\sim 35.0^\circ$~\cite{ANITA:2020gmv} relative to a horizontal plane at the detector position. The shower should have a zenith angle between $\sim (55^\circ-\alpha_C)$ and $\sim (55^\circ+\alpha_C)$ with some extra tolerance of order half to one degree corresponding to emission with an off-axis angle slightly different from $\alpha_C$. 

The particular case studied in \cite{deVries:2019gzs} 
corresponded to a 1 EeV cosmic-ray shower of zenith angle $\psi=55^\circ$, with a shower maximum of $\sim 750~\mathrm{g\,cm^{-2}}$ which is to be compared with a slant depth of the atmosphere of $\sim 1270~\mathrm{g\,cm^{-2}}$ for an altitude of the ice surface above sea level of $\sim 3\,$km. Such a shower will intercept the ice well after shower maximum, when the shower size can be expected to be $\sim 30\%$ of its maximum value\,\footnote{The simulated shower for the mid-TR case of the previous section describes a similar situation.}. Two observers were located at positions such that after reflection they receive the pulses emitted with an off-axis angle  $\alpha=0^\circ$ (aligned with the shower direction and inside the Cherenkov cone) and $\alpha=1.7^\circ$ (outside the Cherenkov cone), so that they are respectively received at $55^\circ$ and $53.3^\circ$ relative to the vertical (see Fig.~\ref{fig:Geometry}). 
A direct comparison performed in \cite{deVries:2019gzs} of the received pulse at $55^\circ$ revealed a change of polarity relative to that observed at $53.3^\circ$.

\begin{figure}[ht]
	\centering
           \subfigure{\includegraphics[width=0.41\linewidth]{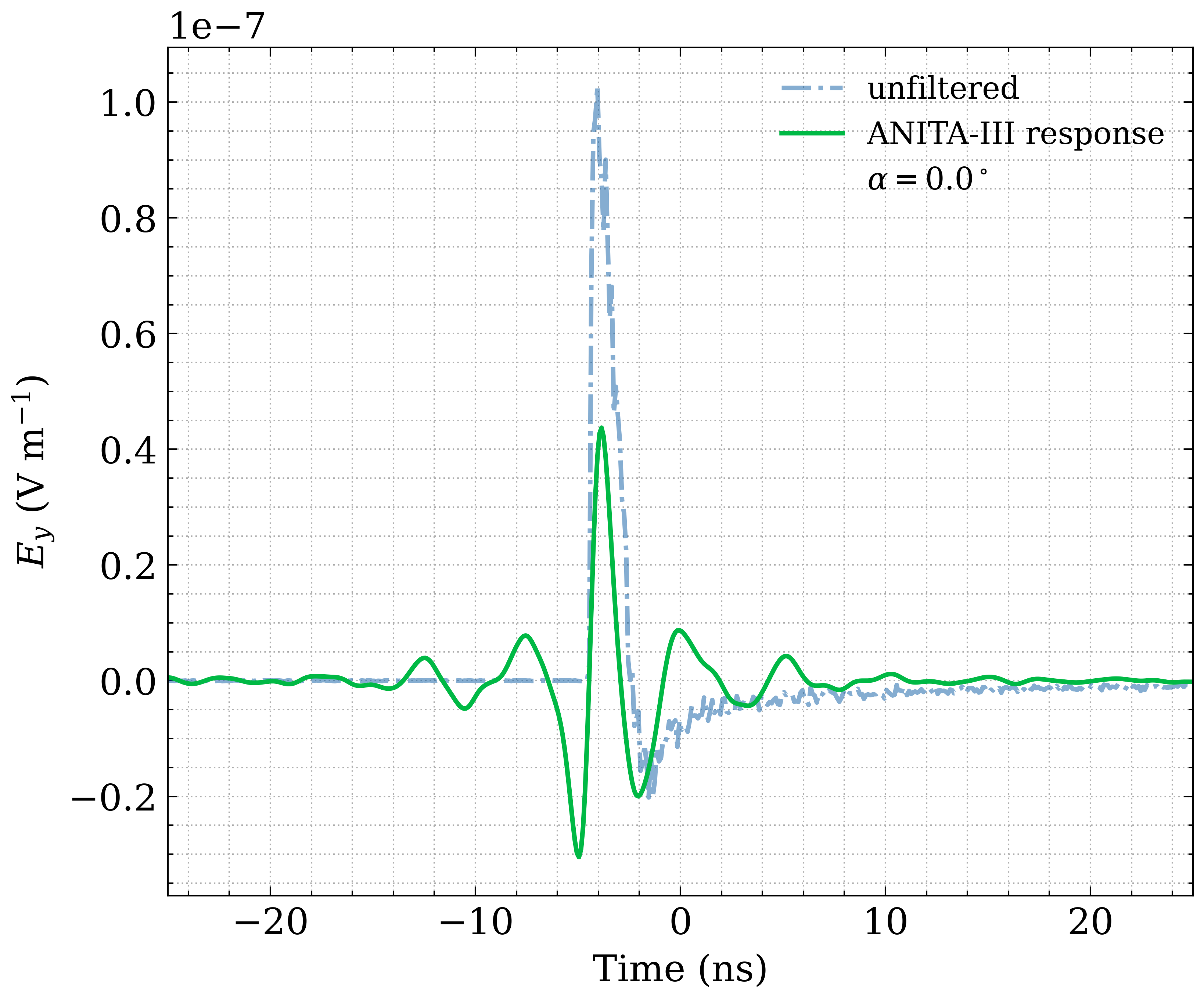}}
    	\subfigure{\includegraphics[width=0.4\linewidth]{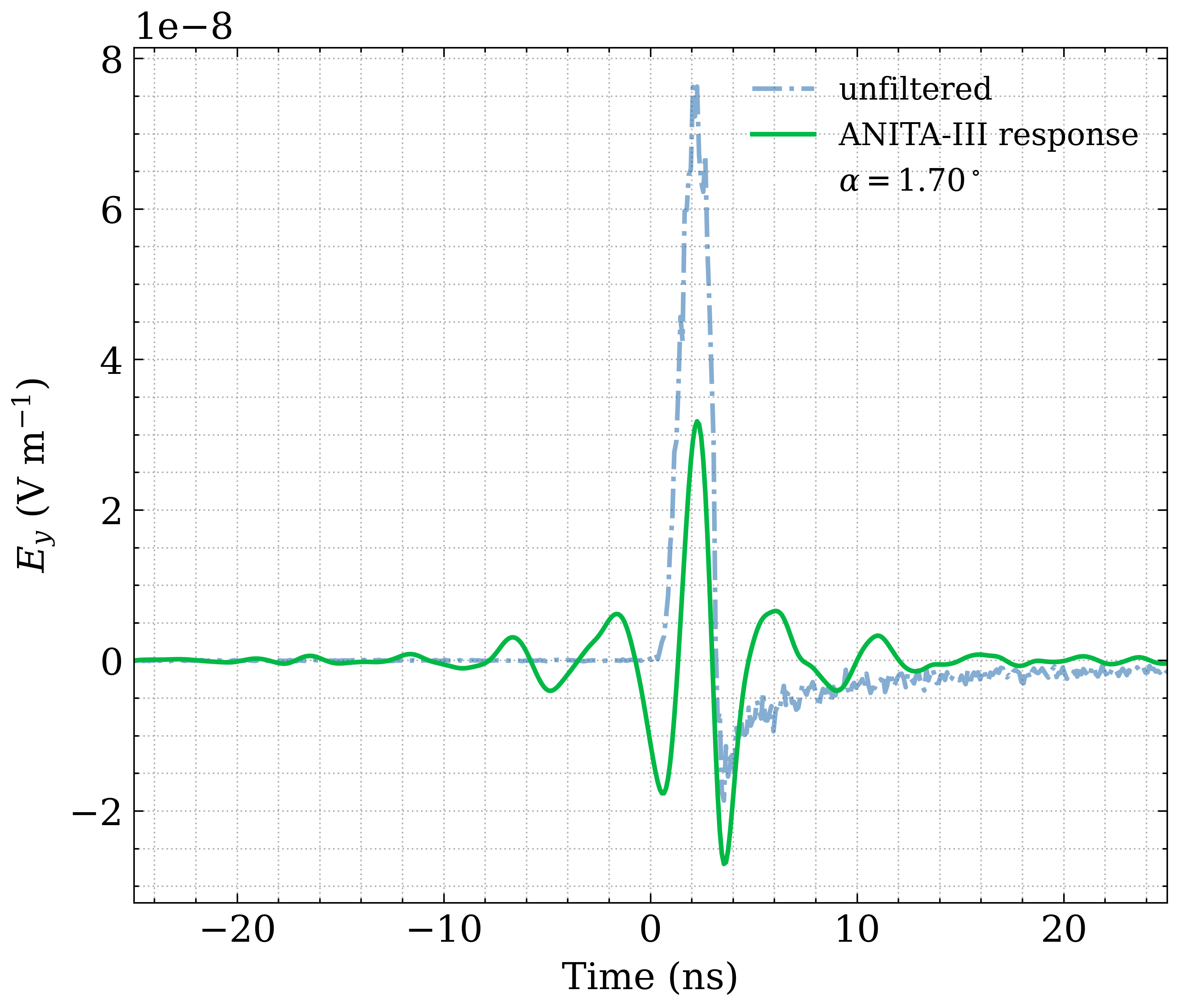}}
\caption{Raw radio pulses (blue dotted-dashed line), obtained with ZHS simulations of a 1 PeV electron shower that intersects the interface at shower maximum (max-TR), as seen by an observer receiving emission with off-axis angle $\alpha=0^\circ$ (left) and $\alpha=1.7^\circ$ (right). The results after a convolution of the pulse with the impulse response of the ANITA III instrument, corrected for time delays, are superimposed (solid green line).}
\label{fig:krijn_Nmax_ZHS_TR_pulses}
\end{figure}

\begin{figure}[ht]
	\centering
            \subfigure{\includegraphics[width=0.41\linewidth]{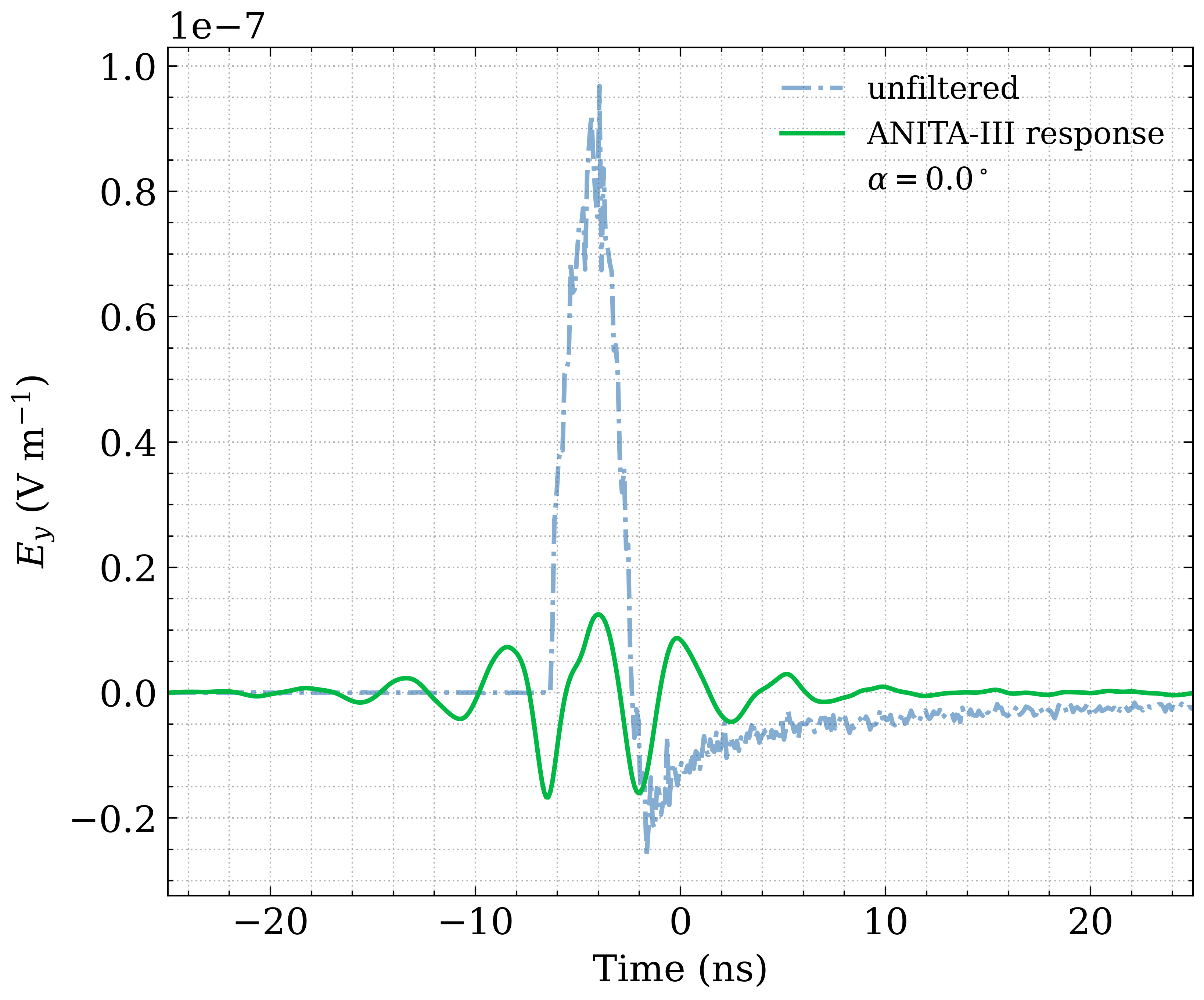}}
    	\subfigure{\includegraphics[width=0.4\linewidth]{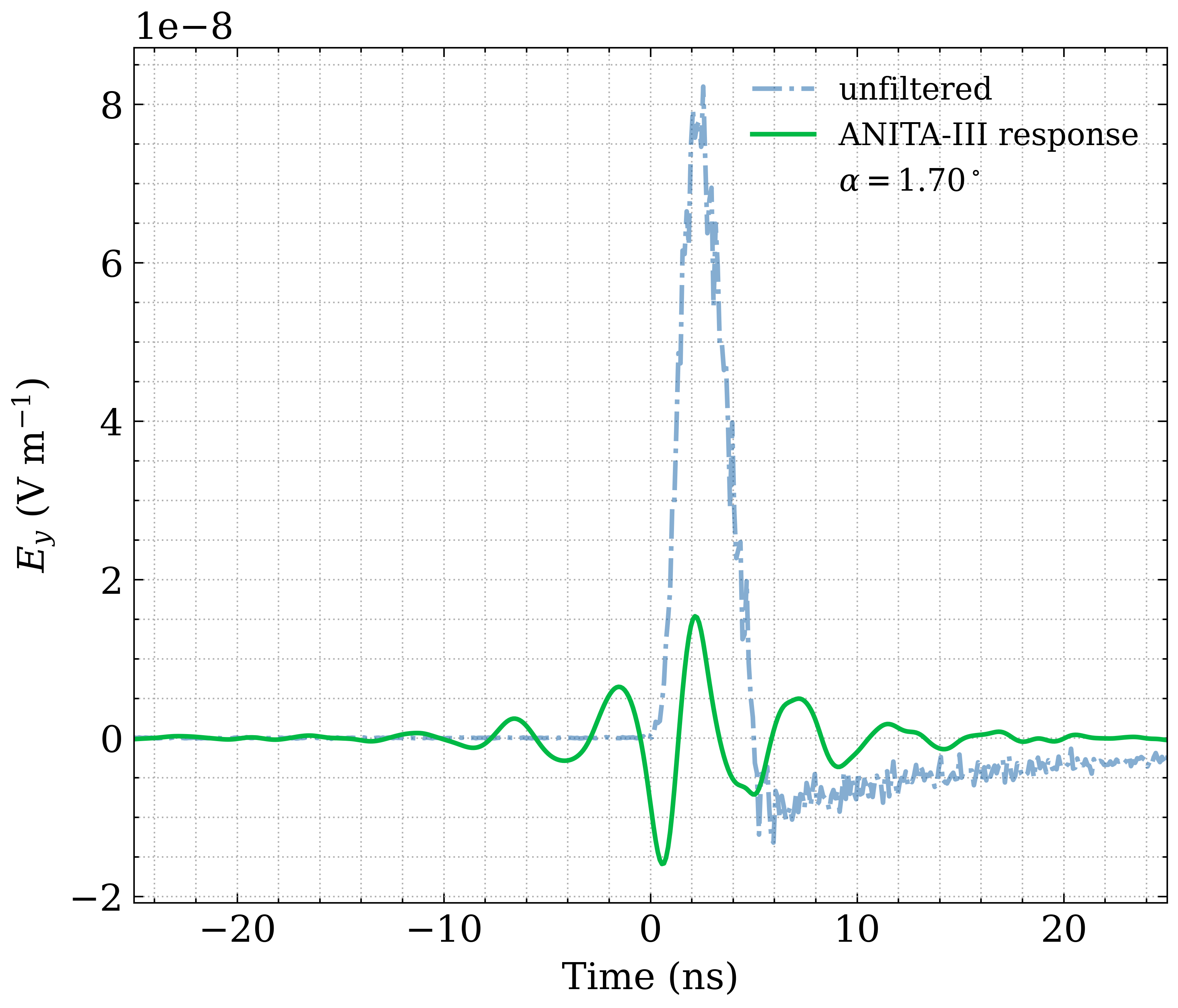}}
\caption{Same as Fig.~\ref{fig:krijn_Nmax_ZHS_TR_pulses} for a shower that crosses the ice-air interface well after shower maximum (mid-TR), such that only half of the particles expected at shower maximum cross the interface.}
\label{fig:krijn_05Nmax_ZHS_TR_pulses}
\end{figure}

In this work, we have obtained radio pulses with the newly developed ZHS code accounting for coherent TR in the maximal and intermediate cases. 
The geometry chosen is the same as that described in Fig.\,\ref{fig:Geometry}, for the same zenith and observation angles as in the case study described above and published in~\cite{deVries:2019gzs}.
We have simulated showers fully contained in the atmosphere and not crossing the ice-air interface (no-TR), with shower maximum before (mid-TR) and at the interface (max-TR). The mid-TR case, when the shower crosses the interface well after reaching shower maximum so that its size is $\sim 50\%$ of its maximum value, is comparable to what could be expected for the cosmic-ray shower studied in~\cite{deVries:2019gzs}. In all cases the pulses are obtained along shower axis $\alpha=0^\circ$ and with an off-axis angle $\alpha=1.7^\circ$, as sketched in Fig.~\ref{fig:Geometry}. The raw, full-band pulses without accounting for the frequency response of the antenna nor the instrument electronics, are shown in blue in in Fig.\,\ref{fig:krijn_Nmax_ZHS_TR_pulses} for the max-TR and in Fig.\,\ref{fig:krijn_05Nmax_ZHS_TR_pulses} for the mid-TR case. 

If a shower is approximated as a 1-dimensional line current i.e. neglecting its lateral spread, it is well known that, for an observer inside the Cherenkov angle, the emission from the end of the shower reaches the observer before the emission from the earlier part~\cite{allan1970frequency}. For the cases with intermediate or maximal TR and $\alpha=0^\circ$, the latest part of the shower then corresponds to the shower intercepting the ground where TR is produced. The converse is true for the observer outside the Cherenkov cone with $\alpha=1.7^\circ$. This picture is preserved under reflection and can be observed provided that the pulse shape is dominated by the development of the shower in the longitudinal direction. 
This effect can be clearly appreciated in Figs.~\ref{fig:krijn_Nmax_ZHS_TR_pulses} and~\ref{fig:krijn_05Nmax_ZHS_TR_pulses}. The effect of TR manifests as an abrupt slope at the onset of the raw pulses on the left plots of both figures and at the end of the first peak in the right plots.
The effect is naturally most apparent in the max-TR case for which the TR component is largest (Fig.~\ref{fig:krijn_Nmax_ZHS_TR_pulses}).
It is also apparent in Figs.\,\ref{fig:krijn_Nmax_ZHS_TR_pulses} and~\ref{fig:krijn_05Nmax_ZHS_TR_pulses} that the polarity does not change when the observer moves from inside ($\alpha<\alpha_C)$ to outside ($\alpha>\alpha_C$) the Cherenkov cone, neither in the case of max-TR nor in the case of mid-TR. These results are consistent with those presented in Section~\ref{sec:pulses} where no polarity changes were identified either, and are in contrast to those presented in~\cite{deVries:2019gzs}. 

In this work we have also studied the influence of the frequency response of the system on the recorded pulses. We have used the measured impulse response of the ANITA III instrument as an example~\cite{rotter2017cosmic}  after correcting for group delay as was done in the published results for ANITA~\cite{ANITA:2018sgj, ANITAprivate_communication}. The results are also shown in green in Figs.~\ref{fig:krijn_Nmax_ZHS_TR_pulses} and~\ref{fig:krijn_05Nmax_ZHS_TR_pulses} along with the raw pulses for comparison. As in the raw pulse case, the convolved pulses for the case of max-TR in Fig.\,\ref{fig:krijn_Nmax_ZHS_TR_pulses} are sharper than the corresponding pulses for mid-TR in Fig.\,\ref{fig:krijn_05Nmax_ZHS_TR_pulses}.

\begin{figure}[ht]
	\centering
    	\subfigure{\includegraphics[width=0.4\linewidth]{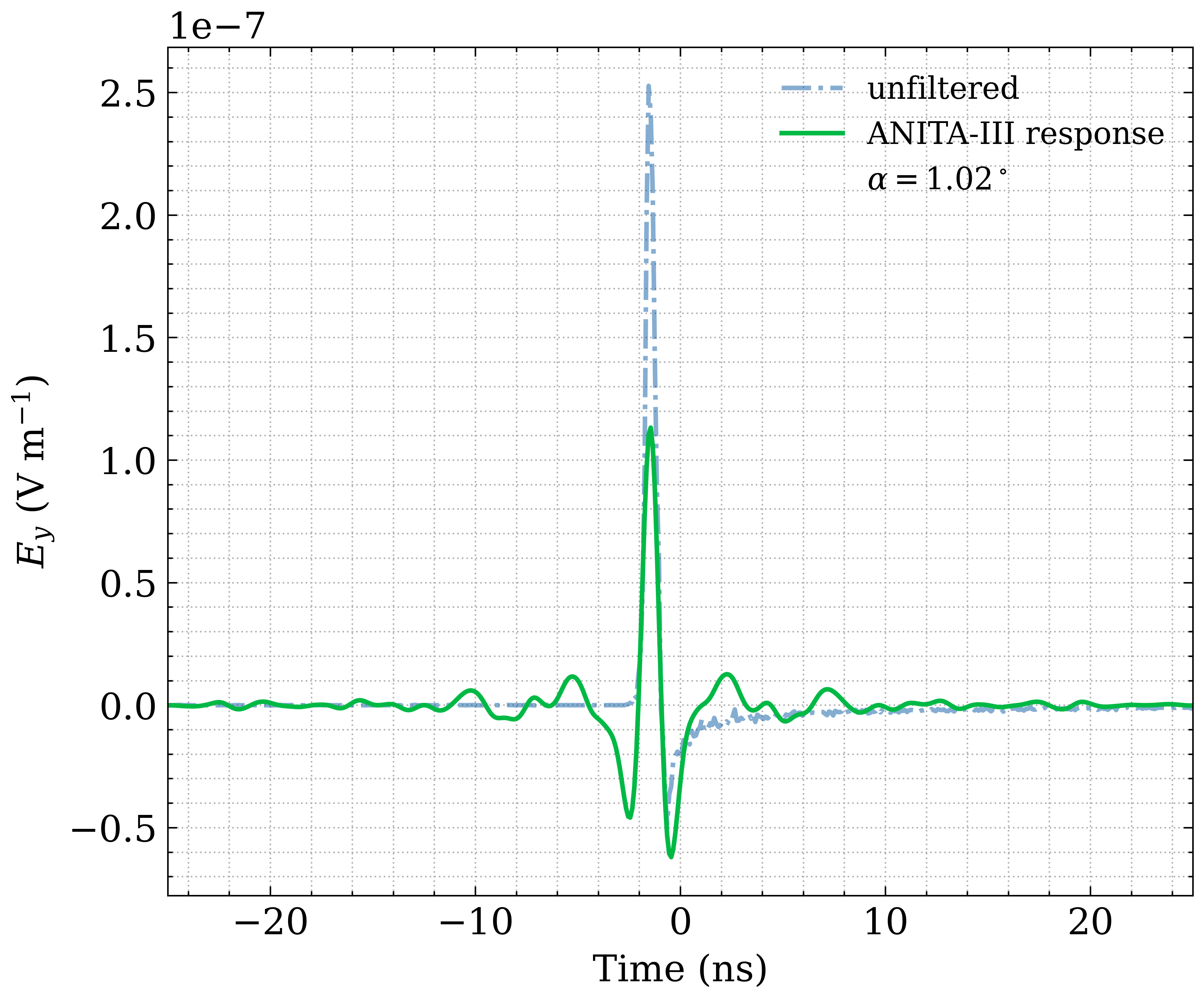}}
    	\subfigure{\includegraphics[width=0.4\linewidth]{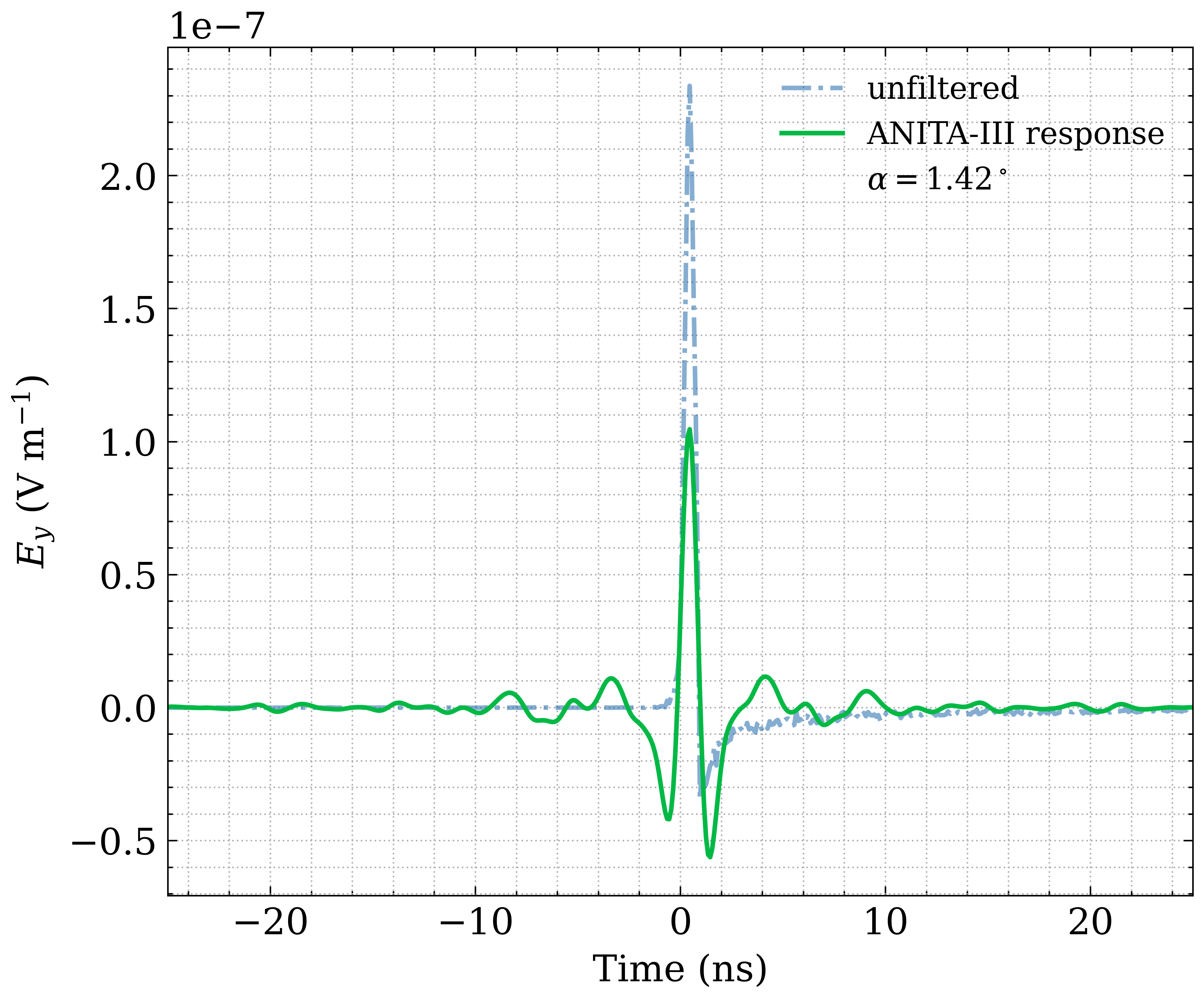}}
\caption{Same as Fig.~\ref{fig:krijn_Nmax_ZHS_TR_pulses} corresponding to the case of max-TR, but for observing angles $\alpha = \alpha_C \pm 0.2^\circ$.}
\label{fig:krijn_Nmax_ZHS_TR_pulses_close_to_cher}
\end{figure}

\begin{figure}[ht]
	\centering
    	\subfigure{\includegraphics[width=0.4\linewidth]{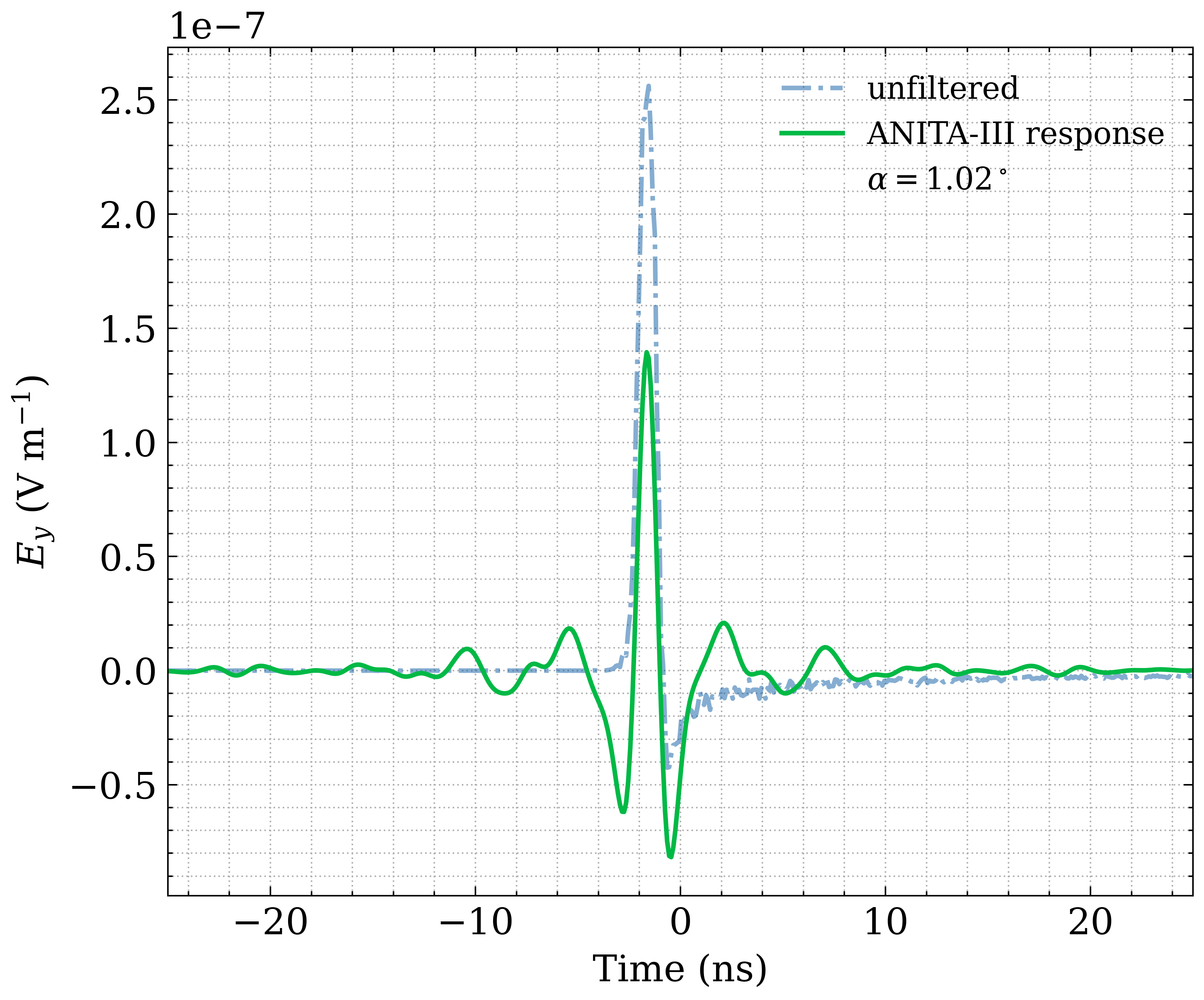}}
    	\subfigure{\includegraphics[width=0.4\linewidth]{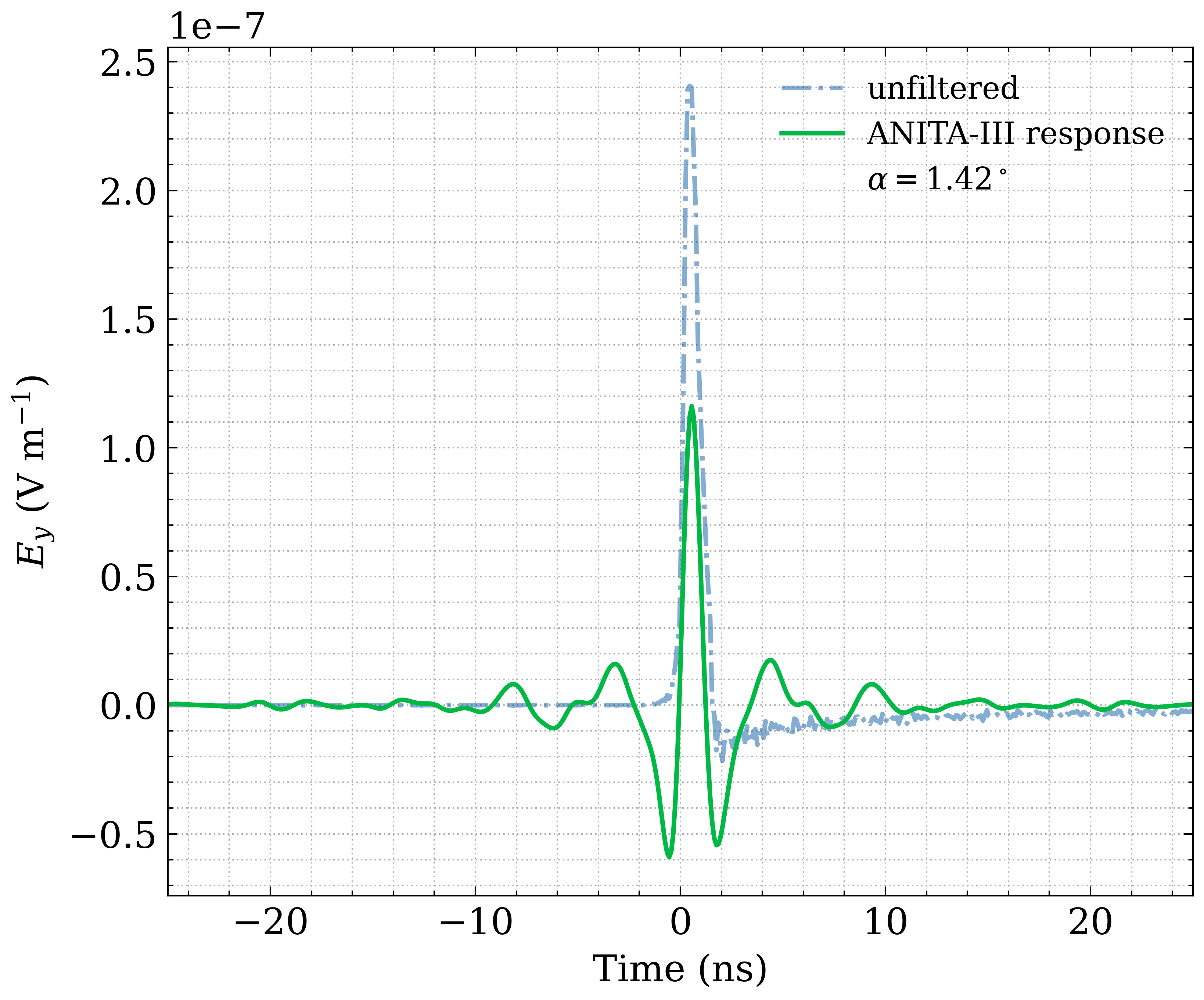}}
\caption{Same as Fig.~\ref{fig:krijn_05Nmax_ZHS_TR_pulses}, corresponding to the case of mid-TR, but for observing angles $\alpha = \alpha_C \pm 0.2^\circ$.}
\label{fig:krijn_05Nmax_ZHS_TR_pulses_close_to_cher}
\end{figure}

In Figs.\,\ref{fig:krijn_Nmax_ZHS_TR_pulses_close_to_cher} and \ref{fig:krijn_05Nmax_ZHS_TR_pulses_close_to_cher} we  display the raw pulses and the pulses after the detector response for the cases mid-TR and max-TR, for off-axis angles close to the Cherenkov angle $\alpha = \alpha_C \pm 0.2^\circ$. Being closer to the Cherenkov direction, the raw pulses are shorter in time compared to those in Figs.\,\ref{fig:krijn_Nmax_ZHS_TR_pulses} and \ref{fig:krijn_05Nmax_ZHS_TR_pulses}, as expected. When convolved with the detector response, the filtered pulses agree closely in shape with the pulses registered in the ANITA III flight for all downward-going cosmic-ray induced showers after being reflected in the ice. The shape of the pulses is also very similar to that of the anomalous event detected by ANITA III in \cite{ANITA:2018sgj}, except for the inverted polarity. 

The results of this work are based on simulations with the ZHS code for showers developing in a homogeneous atmosphere with constant density and refractive index. As shown in Section~\ref{sec:pulses}, it is possible to obtain a good approximation of the pulse observed at high altitude from showers that develop in air and intercept the ice by neglecting the emission from the particles below the surface. This opens the possibility to perform more realistic simulations using the ZHAireS code that deals with the changing density and refractive index of the atmosphere~\cite{Alvarez-Muniz:2011ref}.
A version of ZHAireS to deal with reflection, ZHAireS-Reflex~\cite{alvarez2015simulations}, has been already developed for the analysis of the reflected pulses from downward-going UHECR showers and that have been detected with the ANITA I and III flights~\cite{Schoorlemmer:2015afa}. 
This code removes all particles as soon as they intercept the interface. As a result, it also calculates the effects of transition radiation under this approximation. 

With this version of ZHAireS we simulated 1 EeV proton-induced showers with zenith angle $55^\circ$ and arriving from the magnetic North (see Fig.\,\ref{fig:Geometry}). The showers intercept the air-ice interface at 3 km altitude, typical of the Antarctic continent.
A magnetic field of inclination $-70^\circ$ and magnitude $41.2\,\mu$T was used to match the same conditions in the simulations with the ZHS code. 

\begin{figure}[ht]
	\centering
    	\subfigure{\includegraphics[width=0.52\linewidth]{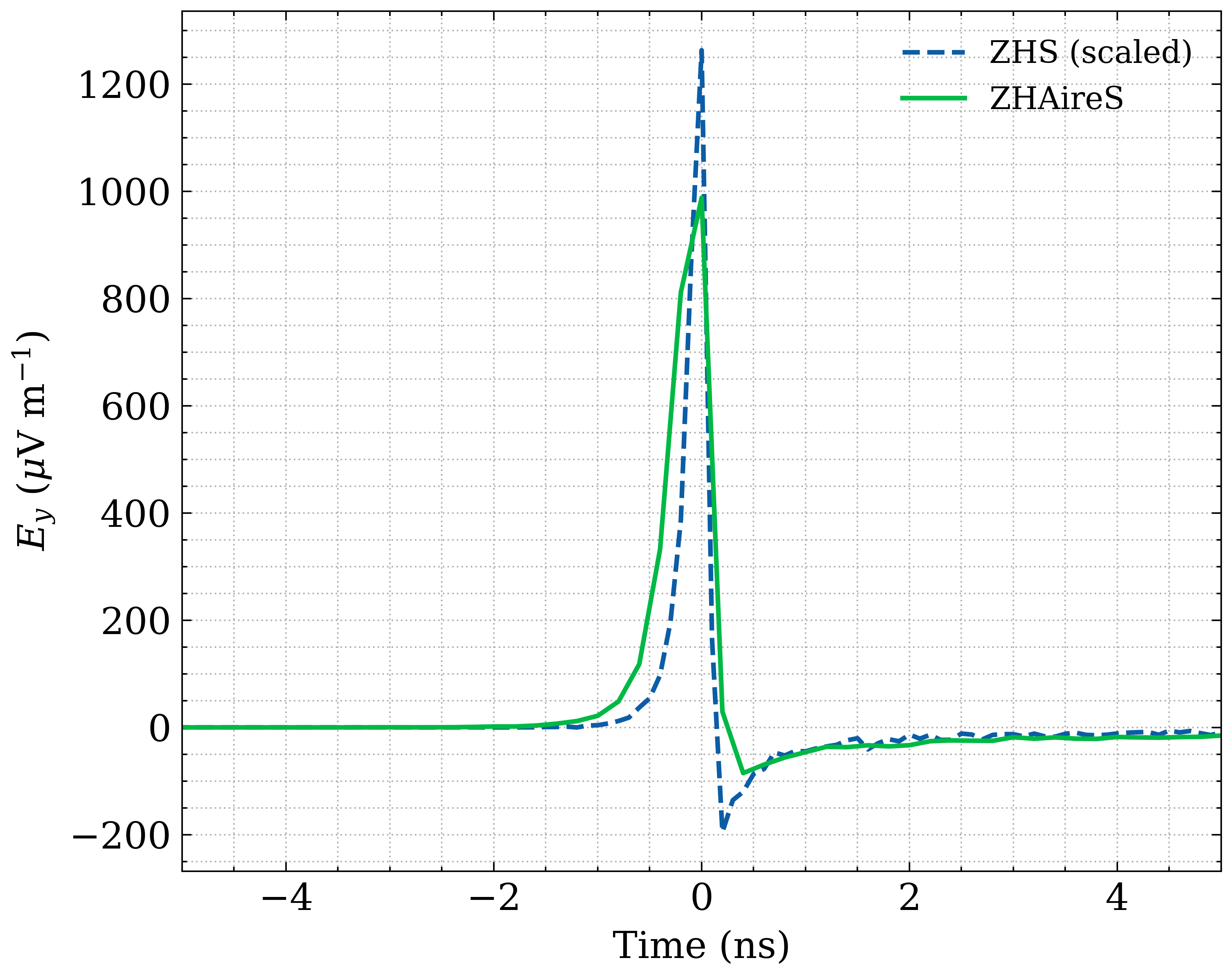}}
\caption{Comparison of the pulses emitted by a shower intersecting the ice at shower maximum simulated with ZHS and ZHAireS. The result of ZHS has been scaled to account for the difference in primary type and energy, electron of 1 PeV in ZHS, proton of 1 EeV in ZHAireS.
}
\label{fig:zhaires_vs_zhs}
\end{figure}
We first show in Fig.\,\ref{fig:zhaires_vs_zhs} a comparison of the previous results obtained with ZHS for a 1 PeV electron-induced shower, and those with ZHAireS for a 1 EeV proton shower. The ZHS simulations were done at the density of air at 3 km altitude.  In both cases the observer is located at 33 km of altitude viewing the shower at an off-axis angle of $\alpha_C=0.78^\circ$, i.e. at the Cherenkov cone after reflection. The ZHS calculation has been scaled up by a factor of 1000 to roughly account for the different sizes of the two showers mainly due to their different energies. The results for the pulses in the time-domain are shown to be in reasonable agreement in both simulations. Most importantly, the polarity of the pulses is the same in both ZHS and ZHAireS simulations and this result gives us confidence that the polarity pattern obtained with ZHS applies also to a shower developing in a more realistic atmosphere. 

\begin{figure}[ht]
	\centering
    	\subfigure[$\alpha=0.00^\circ$]{\includegraphics[width=0.32\linewidth]{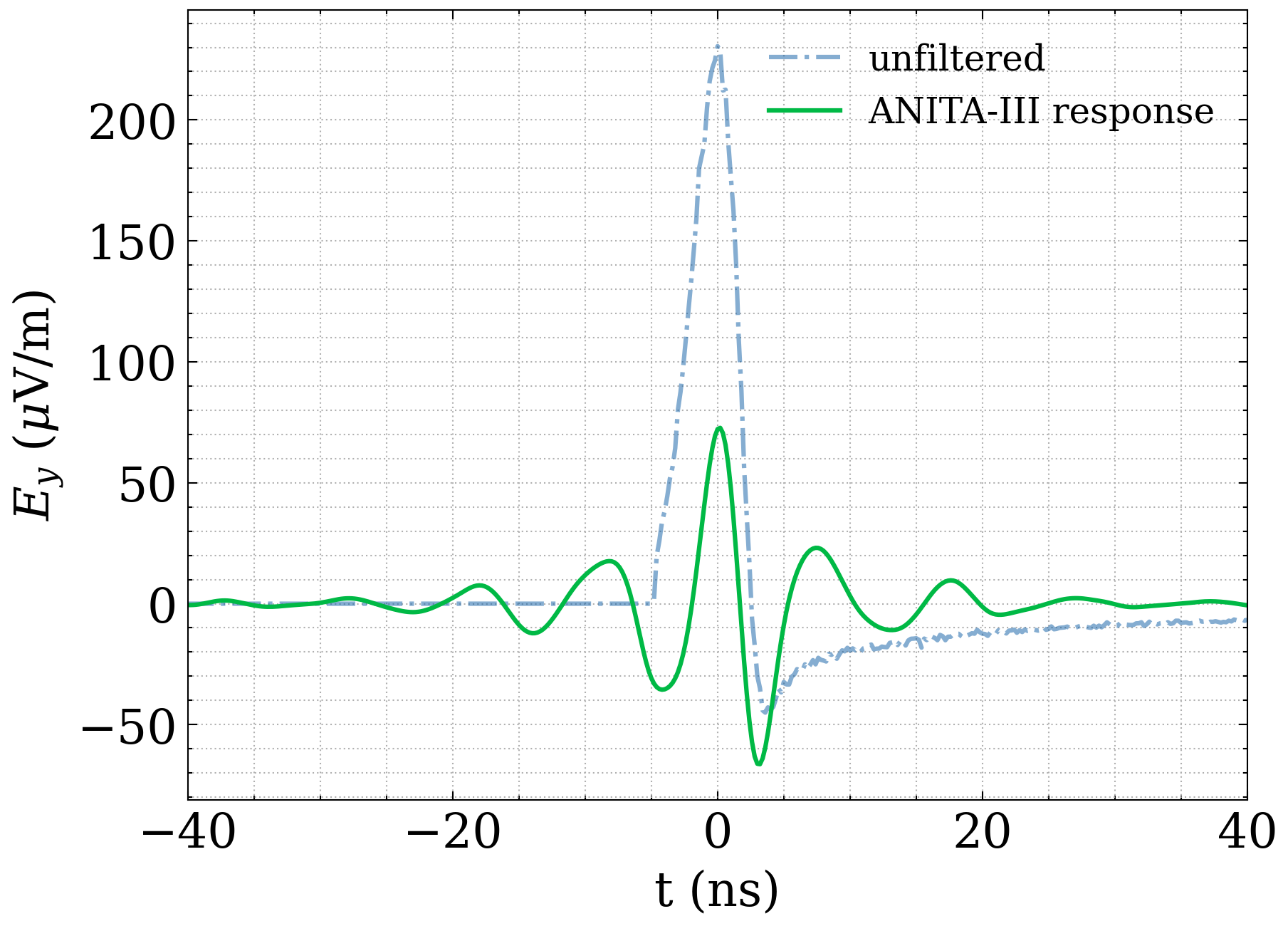}}
    	\subfigure[$\alpha=0.58^\circ$]{\includegraphics[width=0.32\linewidth]{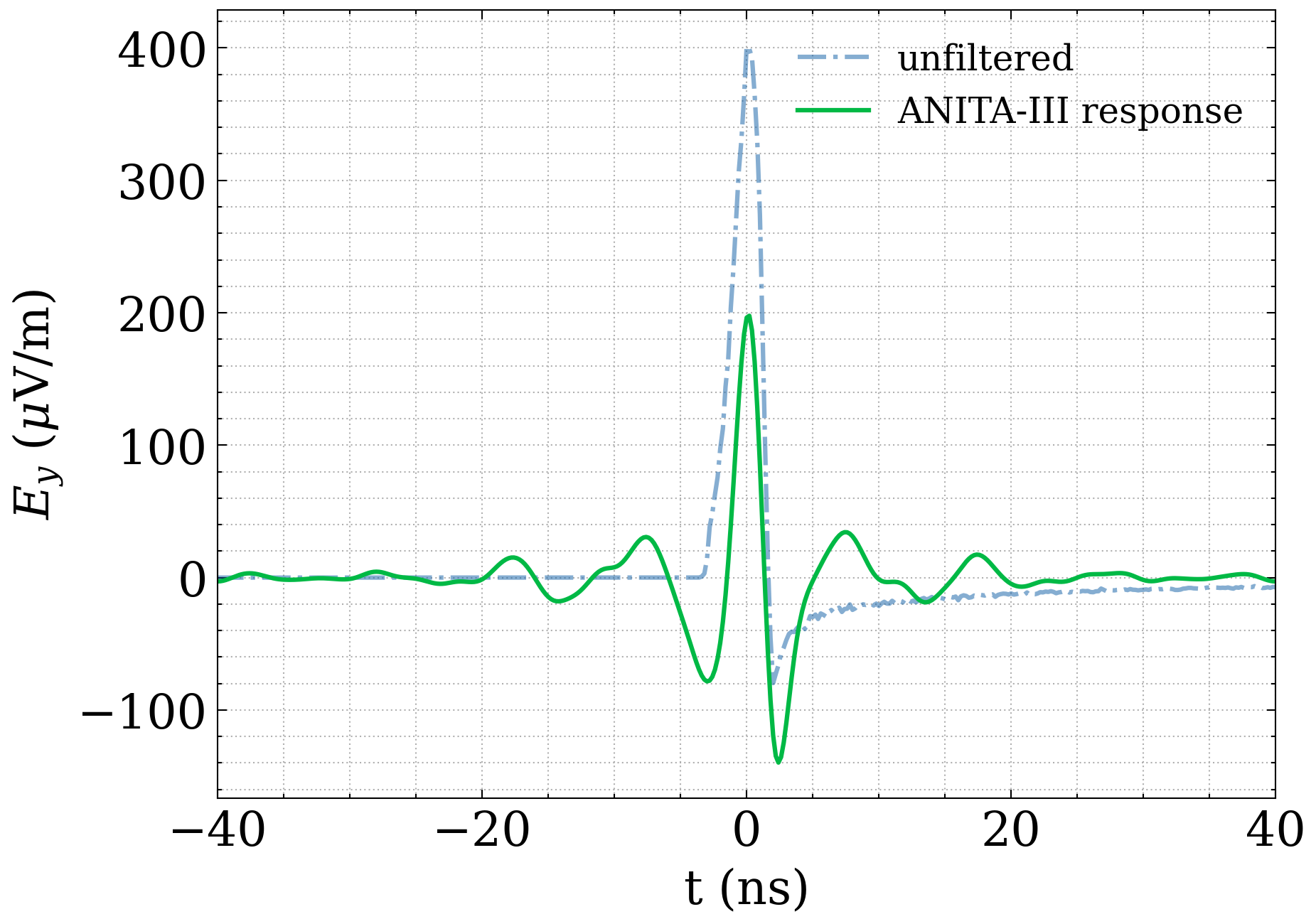}}
    	\subfigure[$\alpha=0.85^\circ$(Cherenkov angle)]{\includegraphics[width=0.32\linewidth]{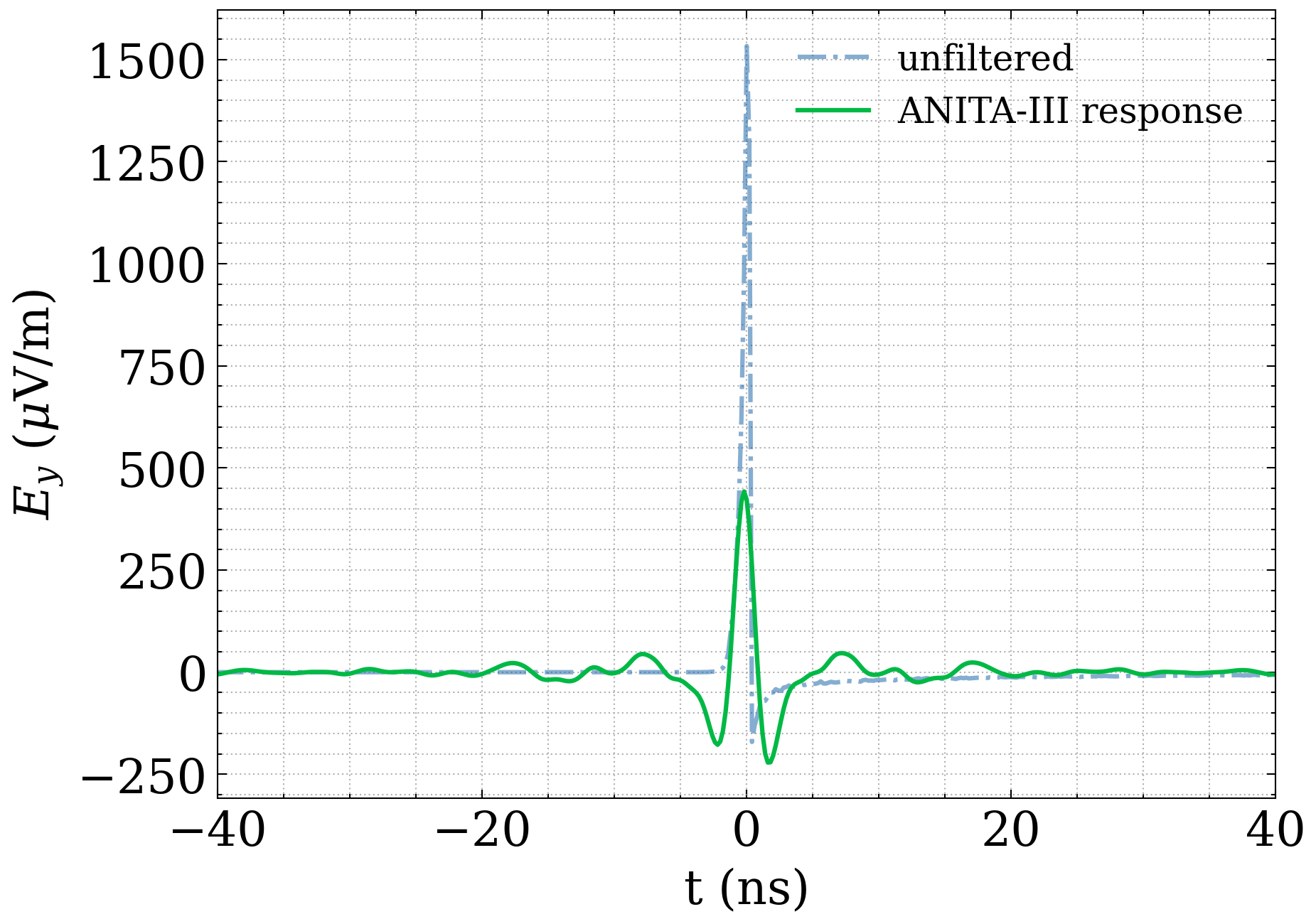}}
            \subfigure[$\alpha=1.11^\circ$]{\includegraphics[width=0.32\linewidth]{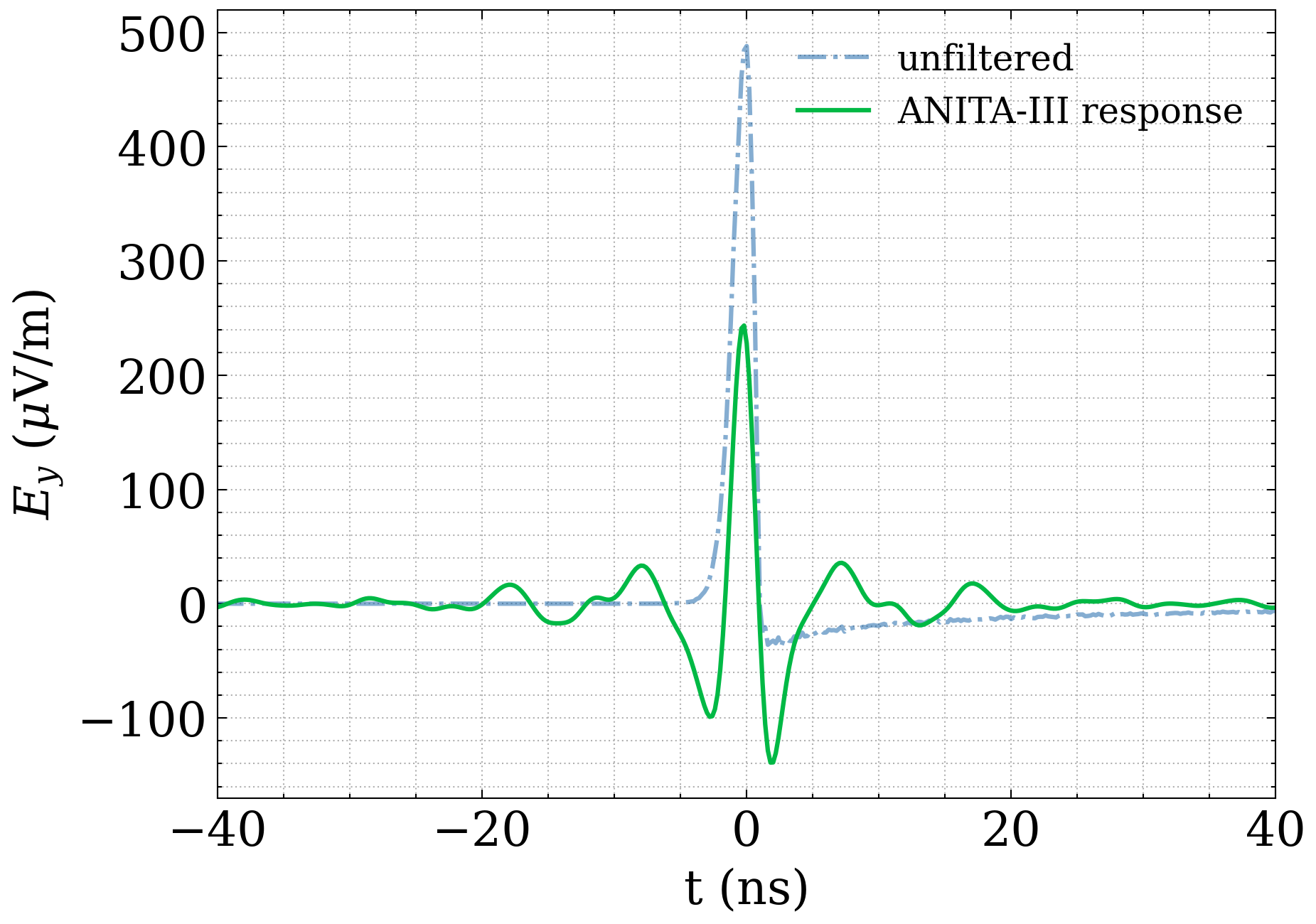}}
    	\subfigure[$\alpha=1.51^\circ$]{\includegraphics[width=0.32\linewidth]{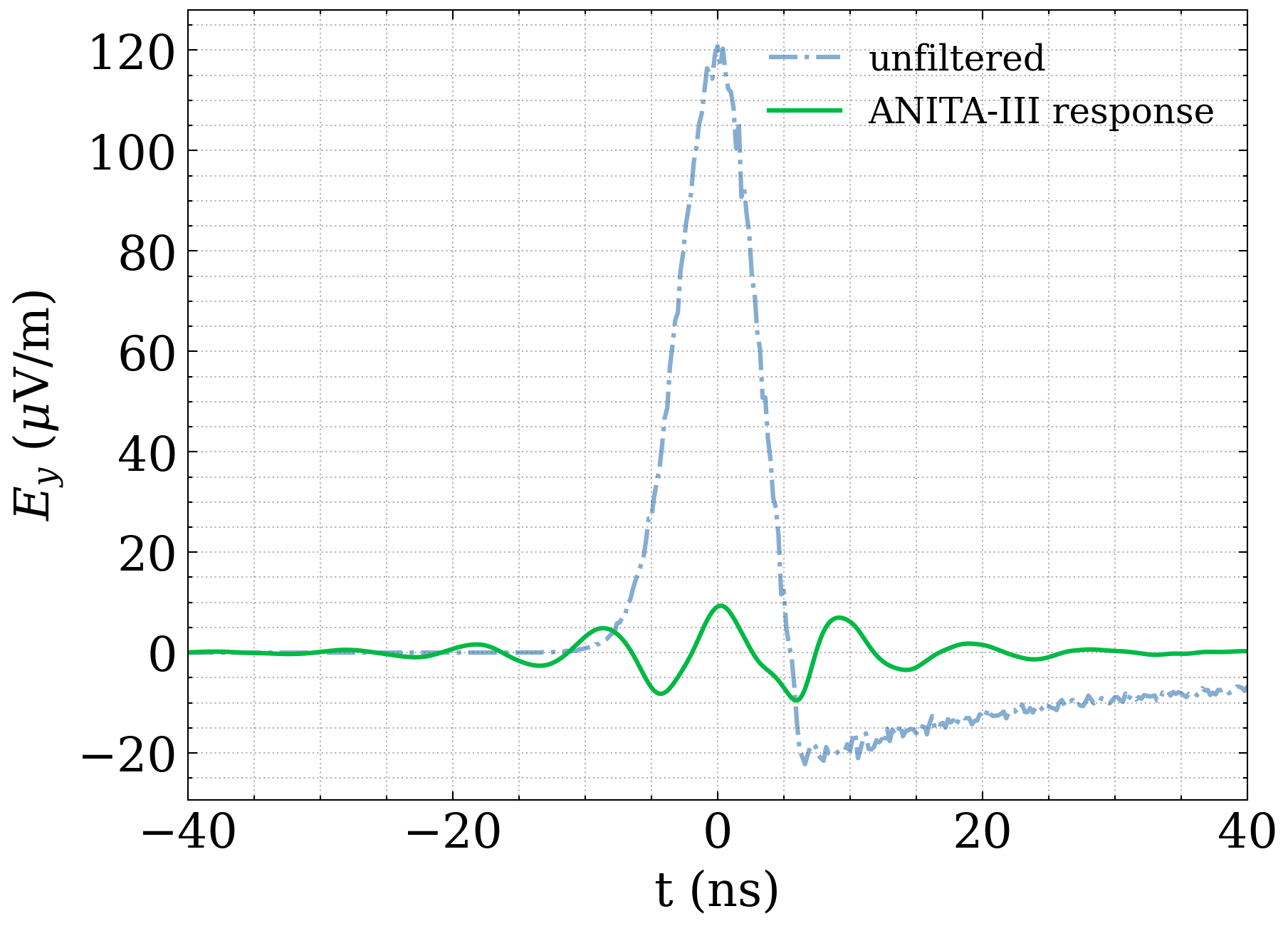}}
    	\subfigure[$\alpha=1.70^\circ$]{\includegraphics[width=0.32\linewidth]{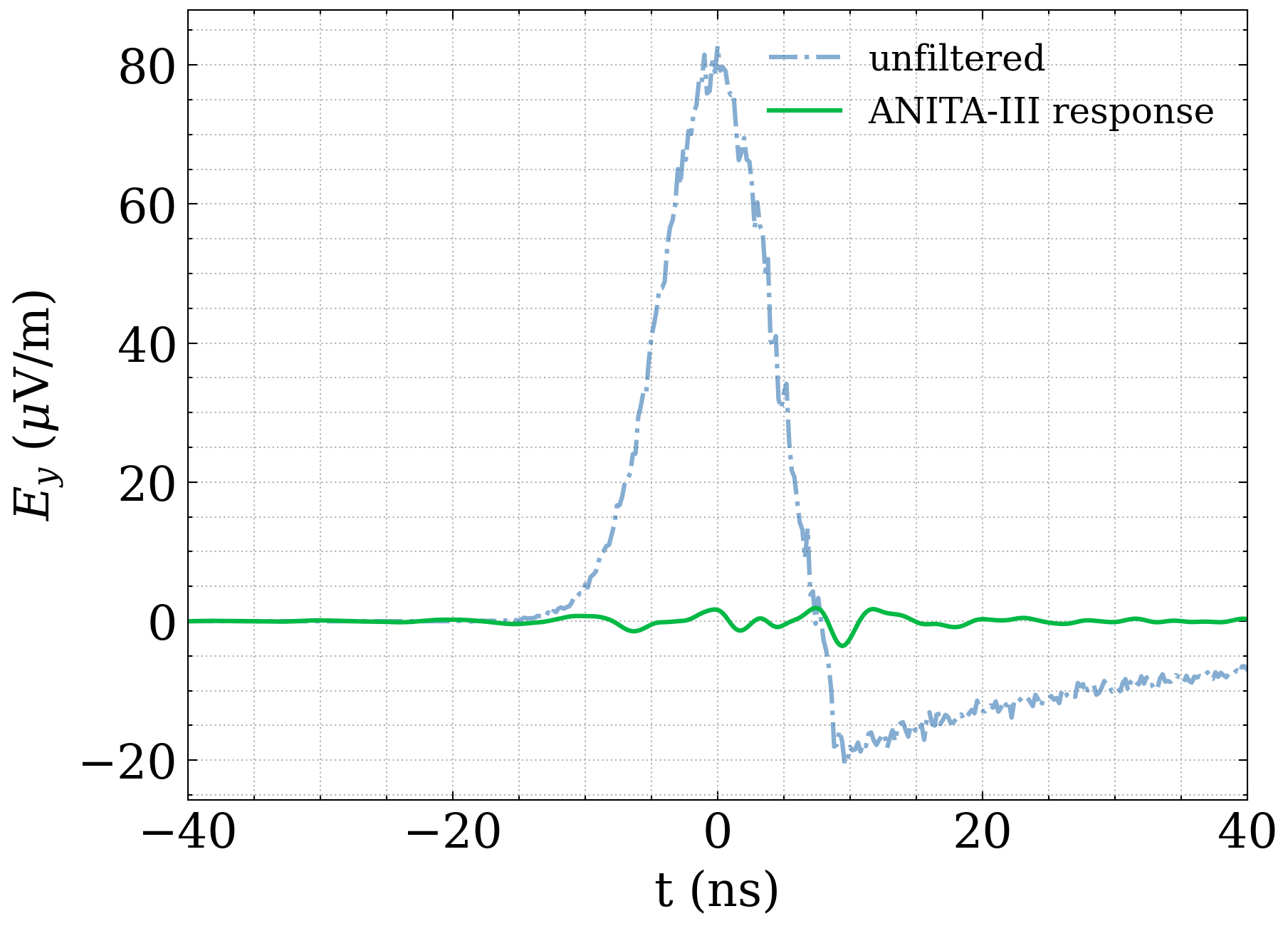}}
\caption{Radio pulses for a 1 EeV proton shower simulated with ZHAireS, for observers at 33 km height and different off-axis angles adopting the geometry in Fig.~\ref{fig:Geometry}. Case (d) corresponds to the Cherenkov direction. The number of particles when the shower reaches ground level (after shower maximum) is of order 30\% of its maximum value (mid-TR case).}
\label{fig:zhaires-reflex55}
\end{figure}

In Fig.~\ref{fig:zhaires-reflex55} we display the pulses for a 1 EeV proton shower with zenith angle $55^\circ$, simulated with ZHAireS-Reflex. The shower intercepts the ice interface at a depth at which the number of particles has dropped to about 30\% its maximum value. The panels correspond to  several off-axis angles, $\alpha=0^\circ$ and $\alpha=1.7^\circ$ as also used in~\cite{deVries:2019gzs}, and also $\alpha=0.85^\circ$, approximately the Cherenkov angle at the density of shower maximum, as well as other intermediate values, both inside the Cherenkov cone, $\alpha=0.58^\circ$, and outside it, $\alpha=1.11^\circ$ and $\alpha=1.51^\circ$. The plots display the raw pulses (dash-dotted blue lines) and convoluted with the approximate impulse response discussed above (green). Some of the expected TR features namely, the abrupt increase or decrease of the pulses, are now less apparent because of the changing refractive index that results in different values of the Cherenkov angle for the maximum of the shower and for the later part when TR is produced. Some qualitative features discussed in the previous sections can be appreciated here. For instance, it is clearly seen how the wider pulses at $\alpha=0^\circ$ and particularly $\alpha=1.70^\circ$ result, after convolution with the detector, in pulses with several oscillations resembling those shown in Fig.~\ref{fig:krijn_05Nmax_ZHS_TR_pulses}. 
Most importantly, again none of the pulses displays polarity inversion relative to pulses with no TR, nor as the off-axis angle of the pulse changes from below to above the Cherenkov angle.

\begin{figure}[ht]\centering
\includegraphics[width=\linewidth]{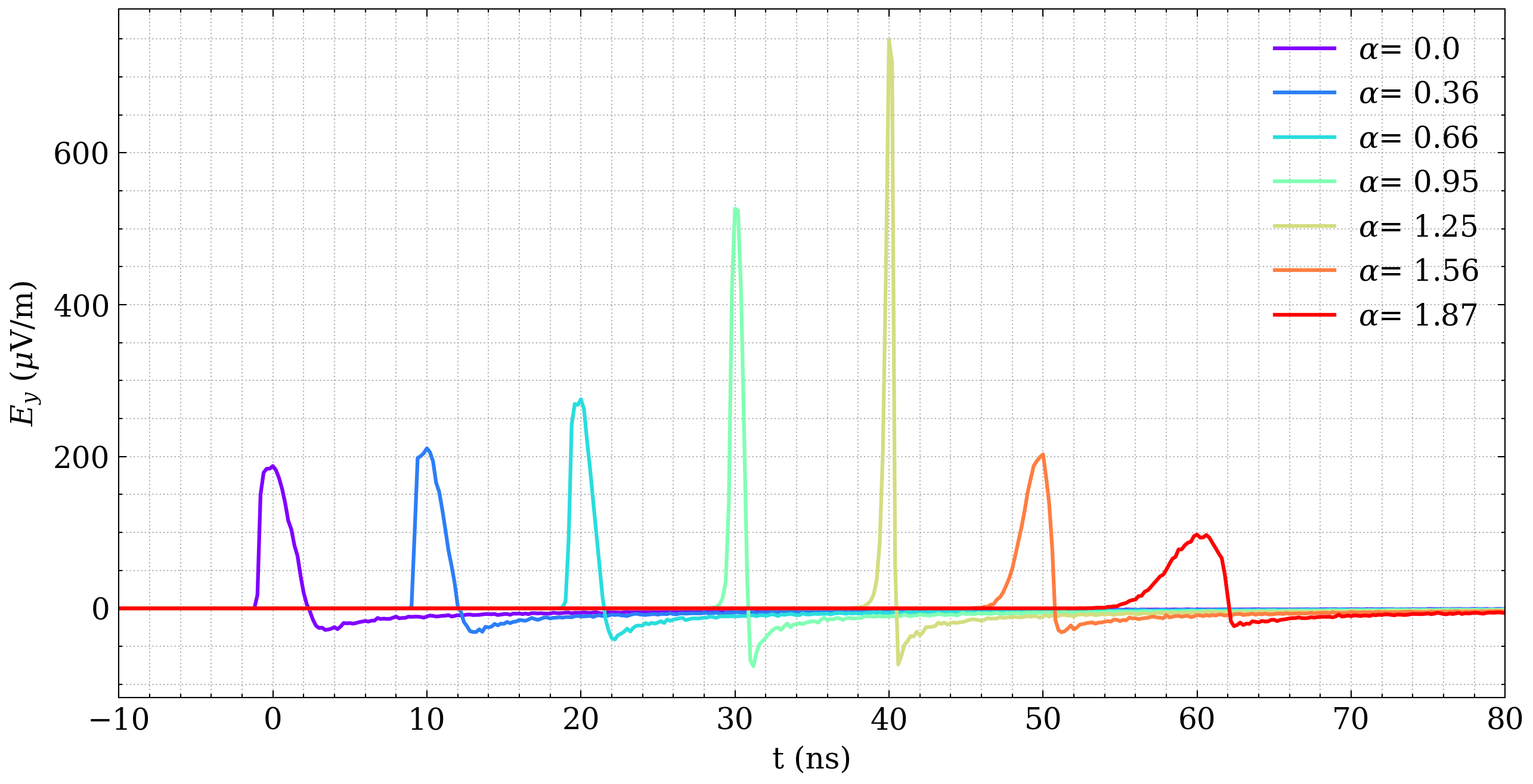}
\caption{Radio pulses obtained using the ZHAireS Monte Carlo for different off-axis angles $\alpha$, for a 1 EeV proton-induced shower with zenith angle $55^\circ$ injected at 10 km of altitude and intersecting ice at 3 km altitude. The number of particles when the shower reaches ground level is close to shower maximum (max-TR case). The observer was placed at 33 km of altitude above the ice cap.} 
\label{fig:zhaires_reflex}
\end{figure}

In Fig.~\ref{fig:zhaires_reflex} we display the raw pulses for a 1 EeV proton shower simulated with ZHAireS-Reflex, with the same zenith angle, $\theta=55^\circ$. In this shower the first interaction point has been forced to take place at a deeper point so that the shower intercepts the interface near shower maximum and the TR effects are more apparent. The plot compares the same off-axis angles plotted in Fig.~\ref{fig:zhaires-reflex55}. The Cherenkov angle, $\alpha_c \sim 1.25^\circ$, is now significantly larger than in the more realistic case of a proton interacting at the top of the atmosphere, because it corresponds to the density at ground level where both shower maximum and TR take place. The effect of transition radiation is now clearly apparent as the steep slope of the pulse, which results in a highly non-symmetric shape. TR takes place in a reduced time compared to the rest of the shower, particularly when the observer is not quite in the Cherenkov angle. 
The abrupt slope is right at the onset of the pulse for off-axis angles below $\alpha_c \sim 1.25^\circ$ because the TR radiation takes place at the end of shower development, while the sequence is inverted within the Cherenkov cone. For angles exceeding $\alpha_c \sim 1.25^\circ$ the steep slope appears on the decay side of the first pulse. Many of the qualitative features discussed in the previous sections can be also appreciated here. In this case with maximal TR, again no polarity inversion can be appreciated as the observing direction moves from inside to outside the Cherenkov cone.

The results using the more realistic ZHAireS-Reflex simulation are compatible with those presented in Section~\ref{sec:TR-ANITA3}, showing again no polarity inversion for pulses observed within and outside the Cherenkov cone and no polarity change compared to pulses that have no TR component and are just reflected on the ice surface. These simulations give further support to the hypothesis that the ZHS code for homogeneous media can indeed be used to simulate TR pulses with sufficient fidelity to test the explanation of the anomalous ANITA events being due to TR from UHECR showers that intercept the ice surface.  

\section{Conclusion}
\label{sec:conclusions}

We have studied the polarity of the coherent radio pulses produced by extensive air showers that intercept a flat ice interface, including transition radiation (TR), as expected for UHECR showers at zenith angles below about $60^\circ$ that develop in Antarctica. The polarity has been studied with the ZHS simulation code for a constant density atmosphere in three different scenarios for maximal TR, for intermediate TR and for no TR expected, respectively, adjusting the depth at which the shower intercepts the ice surface to that of shower maximum, past shower maximum when the size is half of its maximum value, and at a depth well after the full shower has been absorbed in the air. The observer is located at selected  positions such that the emission, making a small off-axis angle relative to the shower axis, reaches the observer after reflection and remains coherent to nearly 1 GHz. The pulses are shown to become slightly sharper in time due to the effect of TR, but it is clearly shown that the polarity of the pulses is the same both for showers with and without TR. 

The polarity has also been studied in the case of showers with TR as the observing location moves from a region inside the Cherenkov angle to a region outside it, showing that the polarity remains the same as the observer moves across the Cherenkov cone. The polarity is preserved in the same way it is preserved for pulses received directly from showers (without the effect of reflection or TR from a flat surface).  

We conclude that the effect of accounting for TR in the radio emission from cosmic-ray showers developing in the atmosphere and intercepting the ice surface, cannot produce the same polarity as that observed in the anomalous ANITA events. This is in contrast to the results of~\cite{deVries:2019gzs}. This explanation is not consistent with the anomalous events detected with the ANITA I and III flights whose nature remains unidentified. This conclusion is not altered when the response of the ANITA I and III instruments is accounted for. 
Therefore, the explanation of the anomalous events still remains an unresolved problem.

\section{Acknowledgements}

This work has received financial support from 
Xunta de Galicia (CIGUS Network of Research Centers),
Consolidaci\'on 2021 GRC GI-2033 ED431C-2021/22 and 2022 ED431F-2022/15;
Ministerio de Ciencia e Innovaci\'on/Agencia Estatal de Investigaci\'on 
PID2019-105544GB-I00, PID2022-140510NB-I00, RED2018-102661-T (RENATA) and PRE2020-092276; and European Union ERDF.

\bibliography{main}

\end{document}